\documentclass[11pt,a4paper]{article}

\usepackage[margin=1in]{geometry}
\usepackage{amsmath,amssymb,amsthm}
\usepackage{graphicx}
\usepackage{booktabs}
\usepackage{array}
\usepackage{xcolor}
\usepackage{hyperref}
\usepackage{caption}
\usepackage{float}
\usepackage{enumitem}
\usepackage{cite}
\usepackage{url}
\usepackage{algorithm}
\usepackage{algpseudocode}

\hypersetup{
  colorlinks=true,
  linkcolor=blue!70!black,
  citecolor=green!50!black,
  urlcolor=blue!70!black,
}

\newcommand{\PLM}{\textsc{PLM-NIDS}}

\newcolumntype{C}[1]{>{\centering\arraybackslash}p{#1}}

\title{\textbf{PLM-NIDS: A Protocol-Language Model for Network Intrusion\\
Detection from Raw Packet Sequences Using RWKV State-Space Models\\
\large (Without Deep Packet Inspection)}}

\author{
  Vivek Kumar Sharma\\
  Palo Alto Networks\\
  \href{mailto:vivsharma@paloaltonetworks.com}{vivsharma@paloaltonetworks.com}
}

\date{May 2026}

\begin{document}
\maketitle

\begin{abstract}
Modern network intrusion detection systems (NIDS) are caught in a structural
contradiction: the protocols carrying the highest threat intelligence are
precisely those encrypted under TLS~1.3 and QUIC, where payload inspection
yields nothing.  We ask a simpler question---\emph{what if the attack
signature is not in the bytes, but in the rhythm?}---and answer it by
treating network flows as a language whose grammar is written entirely in
L3/L4 packet metadata: length, inter-arrival time, TTL, TCP flags, and
hashed port numbers.

We present \PLM{}, which proves three claims in sequence.  \textbf{(1)~The
grammar exists and is learnable}: a RWKV-4 state-space model trained on
344{,}232 unlabelled Monday flows achieves a causal LM validation loss of
0.204, demonstrating that benign traffic has predictable, statistically
consistent structure.  \textbf{(2)~Attacks violate this grammar}: the
per-flow perplexity score cleanly separates benign from attack flows with
PR-AUC\,=\,0.93 using \emph{zero attack labels at training time}.
\textbf{(3)~This separation is architecturally non-trivial}: an LSTM
trained on identical token sequences degenerates to a majority-class
predictor (ROC-AUC\,$\approx$\,0.50, F1\,=\,0.91 by always predicting
``attack''), proving that RWKV's causal pre-training provides an inductive
bias unavailable to direct classifiers.  Supervised fine-tuning further
raises PR-AUC to 0.94 and ROC-AUC to 0.75, with a precision of 97.7\%
at the calibrated operating threshold.

The RWKV backbone's O(T) recurrent inference enables per-packet streaming
without flow buffering, making \PLM{} operationally viable at line rate.
Because it reads only IP/TCP/UDP headers, it is inherently encryption-agnostic:
TLS~1.3, QUIC, and future encrypted protocols are handled transparently.
\end{abstract}


\tableofcontents
\newpage

\section{Introduction}
\label{sec:intro}

\subsection{The Problem: DPI Is Structurally Broken for Encrypted Traffic}

Deep Packet Inspection has served as the backbone of commercial NIDS for
two decades.  But its fundamental assumption---that payloads carry the
distinguishing signature of an attack---is increasingly invalid.  TLS~1.3
now accounts for over 90\% of HTTPS connections~\cite{googletransparency2024},
and QUIC is displacing TCP for high-throughput applications.  When payloads
are encrypted, DPI-based signatures simply do not fire.  The attacker's
traffic looks identical to the defender's legitimate data at the byte level.

The standard ML response---extract statistical features with tools like
CICFlowMeter~\cite{cicflowmeter2017} and train a classifier---suffers from
the same underlying problem.  CICFlowMeter performs application-layer
parsing to compute its 80+ features; it is a DPI tool wrapped in a Python
interface.  Systems built on it inherit its brittleness: they break on
protocol version changes, require maintenance as traffic evolves, and
cannot extract meaningful features from truly opaque encrypted streams.

\subsection{Our Insight: Protocol Grammar Lives in the Metadata}

Our central observation is that \emph{the behavioural signature of a
protocol---and of an attack---is encoded in packet metadata, not in
payload bytes.}  Consider:
\begin{itemize}[nosep]
  \item A TCP handshake produces a stereotyped SYN\,$\to$\,SYN-ACK\,$\to$\,ACK
        timing and flag sequence.
  \item A bulk HTTP/2 download generates large forward packets with small
        backward ACKs in a characteristic rhythm.
  \item A DoS Slowloris attack holds connections open with tiny, infrequent
        packets---a pattern visible purely in inter-arrival times and lengths.
  \item An SSH brute-force produces many short, symmetrical flows to
        port~22 with a metronomic cadence.
\end{itemize}

None of these signatures require reading a single application byte.  The
\emph{grammar} of normal and abnormal behaviour is inscribed in the timing,
sizing, and flag patterns of packets.

\subsection{Our Approach: Language Modelling on Protocol Tokens}

We formalise this insight as a language modelling problem.  Each packet
is converted to a 9-token tuple encoding direction, binned length,
binned inter-arrival time, binned TTL, L4 protocol, and hashed
source/destination ports---all derivable from standard NIC headers.
A network flow becomes a token sequence, and the problem of modelling
normal traffic becomes the problem of training a language model on these
sequences.

We choose RWKV-4~\cite{peng2023rwkv} as the backbone for two reasons.
First, its O(T) recurrent computation versus the O(T\textsuperscript{2})
attention of Transformers~\cite{vaswani2017attention} is critical for
long flows (DoS attacks can exceed 1{,}000 tokens per flow).  Second,
its streaming \texttt{step()} API processes each packet the instant it
arrives, updating a compact per-flow hidden state without requiring the
complete flow to be buffered---a fundamental requirement for line-rate
detection.

\subsection{What We Claim and Prove}

This paper makes and empirically validates three claims on 2.7~million
real-world flows from CIC-IDS-2017~\cite{sharafaldin2018cicids}:

\begin{enumerate}[label=\textbf{C\arabic*.}, leftmargin=*]
  \item \textbf{Benign traffic grammar is learnable from unlabelled data.}
        A RWKV causal LM trained on benign-only flows converges to
        validation loss~=~0.204 within 10~epochs, demonstrating statistically
        consistent structure in normal enterprise traffic.

  \item \textbf{Attacks violate this grammar in a detectable way.}
        Per-flow perplexity scores---computed without any attack
        labels---achieve PR-AUC\,=\,0.93, with precision\,=\,0.977
        at the calibrated operating threshold.

  \item \textbf{The RWKV architecture is essential; the tokens alone are
        insufficient.}
        An LSTM trained on the same token sequences achieves
        ROC-AUC\,=\,0.50 (equivalent to random), exposing that its
        apparent F1\,=\,0.91 is an artefact of majority-class prediction.
        RWKV's causal pre-training provides an inductive bias---an anchor
        in ``what normal looks like''---that prevents this collapse.
\end{enumerate}

\paragraph{Contributions.}
\begin{enumerate}[leftmargin=*, label=(\arabic*)]
  \item A DPI-free, 9-token-per-packet tokenisation scheme derived
        purely from L3/L4 headers, forming a compact closed vocabulary
        of 227 tokens.
  \item A two-phase training strategy: unsupervised causal LM
        pre-training on benign-only traffic (Phase~1), followed by
        optional supervised fine-tuning (Phase~2).
  \item Empirical proof that RWKV's causal pre-training is
        \emph{architecturally necessary}---not just beneficial---through
        the LSTM collapse experiment.
  \item A streaming per-flow inference pipeline with O(1) memory
        per active flow and O(T) compute per flow, enabling deployment
        at network line rate.
\end{enumerate}

\section{Background and Related Work}
\label{sec:background}

\subsection{Signature-Based NIDS and the DPI Bottleneck}

Systems such as Snort~\cite{roesch1999snort} and Suricata maintain
curated rule databases and match packet payloads against known attack
signatures.  They achieve high precision on known threats but are blind
to zero-day attacks by construction, and fail completely when payloads
are encrypted.  Their rule bases require continuous expert maintenance
as protocols evolve---a cost that scales with the diversity of the
network.

\subsection{Feature-Engineered ML: DPI in Disguise}

The dominant ML approach to NIDS uses CICFlowMeter or equivalent
tools to extract 80+ per-flow statistical features (packet lengths,
inter-arrival times, flag counts, etc.) from raw
PCAPs~\cite{cicflowmeter2017,sharafaldin2018cicids}.  Random Forest and
XGBoost classifiers trained on these features achieve strong results
on standard benchmarks.  However, this approach has three fundamental
limitations: (i)~features require a complete flow before classification,
precluding early detection; (ii)~CICFlowMeter performs application-layer
parsing internally, making these ``DPI-free'' claims misleading; and
(iii)~the feature set requires re-engineering as protocols change.

\subsection{Deep Learning on Raw Traffic}

CNNs operating on raw packet bytes~\cite{wang2017malware} and LSTMs
on packet-length sequences~\cite{liu2019lstm} have been proposed as
parser-free alternatives.  Both approaches suffer from the same
fundamental limitation: they depend on payload bytes that are unavailable
under encryption.  Header-restricted approaches---such as traffic
fingerprinting from timing and size side-channels---are closer in spirit
to our work but do not frame the problem as language modelling.

\subsection{Transformer-Based Pre-training for Network Security}

ET-BERT~\cite{lin2022etbert} applies BERT pre-training to the first
bytes of TLS payloads.  NetBERT~\cite{netbert2020} pre-trains a
Transformer on hex-encoded packet bytes.  Both achieve strong results
but require payload access, which is precisely what is unavailable under
TLS~1.3.  FlowBERT~\cite{flowbert2021} applies Transformer pre-training
to CICFlowMeter features, inheriting the DPI dependency.  \PLM{} differs
fundamentally: it operates on L3/L4 header tokens only, with no access
to any payload byte, in any mode.

\subsection{State-Space Models: O(T) Sequential Modelling}

RWKV~\cite{peng2023rwkv} reformulates attention as a linear recurrence,
achieving Transformer-quality language modelling at O(T) inference cost
and O(1) streaming memory.  Mamba~\cite{gu2023mamba} introduces
input-dependent state transitions.  Both have matched Transformer
performance on NLP benchmarks.  Neither has been applied to network
traffic modelling.  To the best of our knowledge, \PLM{} is the first
work to apply an SSM backbone as a protocol-language model for NIDS,
with streaming per-flow state and no DPI.

\section{Method}
\label{sec:method}

The design of \PLM{} follows directly from our three claims.  To prove
C1 (grammar is learnable), we need a tokenisation that preserves
behavioural signal without payload.  To prove C2 (attacks violate
grammar), we need a model that can score sequence likelihood.  To prove
C3 (architecture matters), we need to run the same tokens through an
LSTM as a controlled comparison.

\subsection{DPI-Free Tokenisation}
\label{sec:tokenisation}

\paragraph{Why these features?}
Each token field is chosen because it carries behavioural information
about the protocol interaction, not about the application payload.
Direction captures the request/response asymmetry of protocols.
Packet length encodes application-layer framing patterns without revealing
content.  Inter-arrival time captures timing rhythms that differ
fundamentally between normal traffic, brute-force scanning, and DoS attacks.
TTL provides a coarse path-distance signal.  Hashed ports provide
protocol-type context without requiring a port-number lookup table.
Together, these nine values per packet characterise \emph{what} is
happening without revealing \emph{what is being communicated}.

\paragraph{Flow keying.}
Packets are grouped into bidirectional flows using the normalised
5-tuple $(\min(A,B), \max(A,B), \text{proto})$ where $A=(src,sport)$
and $B=(dst,dport)$, ensuring packets in both directions share one flow
record.  Flows are evicted on TCP FIN/RST or after 120\,s of inactivity.

\paragraph{Per-packet tokenisation.}
From each packet we derive exactly 9 integer token IDs using fields
available without any payload inspection:

\vspace{0.3em}
\begin{algorithmic}[1]
\Procedure{Tokenise}{packet $p$, flow state $f$}
  \State $\text{dir} \gets \texttt{DIR\_C2S}$ if $p.src = f.origin$ else $\texttt{DIR\_S2C}$
  \State $\text{len} \gets \text{Bin}(\log(1+|p|),\,\mathbf{e}_\ell)$ \Comment{binned log-length}
  \State $\Delta t \gets p.ts - f.last\_ts$;\quad $\text{iat} \gets \text{Bin}(\log(1+\Delta t),\,\mathbf{e}_t)$
  \State $\text{ttl} \gets \text{Bin}(p.ip.ttl,\,\mathbf{e}_h)$
  \State $\text{proto} \gets \textsc{TCP} \mid \textsc{UDP} \mid \textsc{ICMP}$
  \State $\text{sport} \gets \text{offset} + (p.sport \bmod B_p)$
  \State $\text{dport} \gets \text{offset} + (p.dport \bmod B_p)$
  \State $\text{flags} \gets \text{dominant TCP flag token}$
  \State \Return $[\text{dir, len, iat, ttl, proto, sport, dport, flags}, \langle\text{SEP}\rangle]$
\EndProcedure
\end{algorithmic}
\vspace{0.3em}

Bin edges $\mathbf{e}_\ell, \mathbf{e}_t, \mathbf{e}_h$ are fitted
once on training packets using percentile spacing, capturing the
empirical distribution of each field.

\begin{figure}[htbp]
  \centering
  \includegraphics[width=0.65\linewidth]{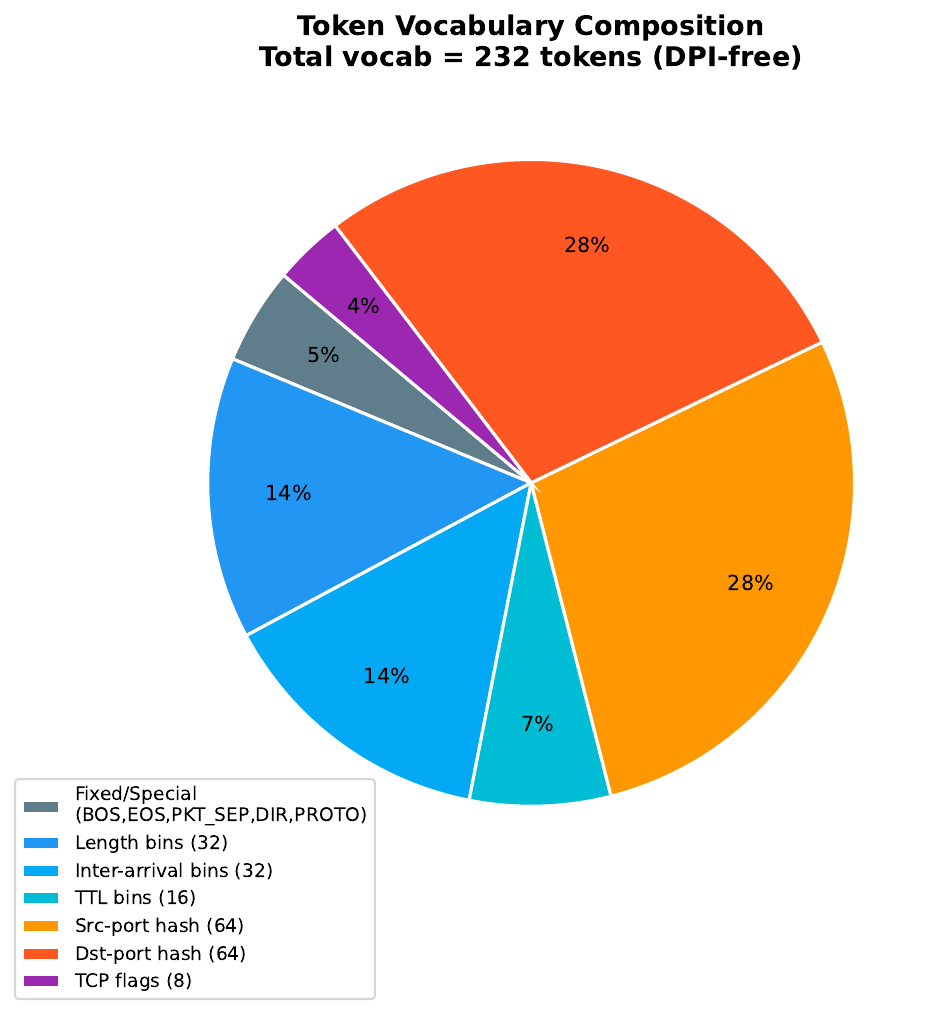}
  \caption{Token vocabulary composition.  All 227 tokens are derived
    exclusively from L3/L4 header fields.  Src-port and dst-port hash
    buckets are the largest groups (28\% each), reflecting the importance
    of port patterns in protocol identification without explicit port
    number lookup.}
  \label{fig:vocab}
\end{figure}

\paragraph{Vocabulary.}
The complete vocabulary has 227 tokens (Figure~\ref{fig:vocab}): 5~special
tokens (\texttt{PAD, BOS, EOS, SEP, UNK}), 6~structural tokens (protocol
and direction), 32~length bins, 32~IAT bins, 16~TTL bins, 64~source-port
and 64~destination-port hash buckets, and 8~TCP-flag tokens.
\emph{No payload byte appears anywhere in this vocabulary.}

\begin{figure}[htbp]
  \centering
  \includegraphics[width=\linewidth]{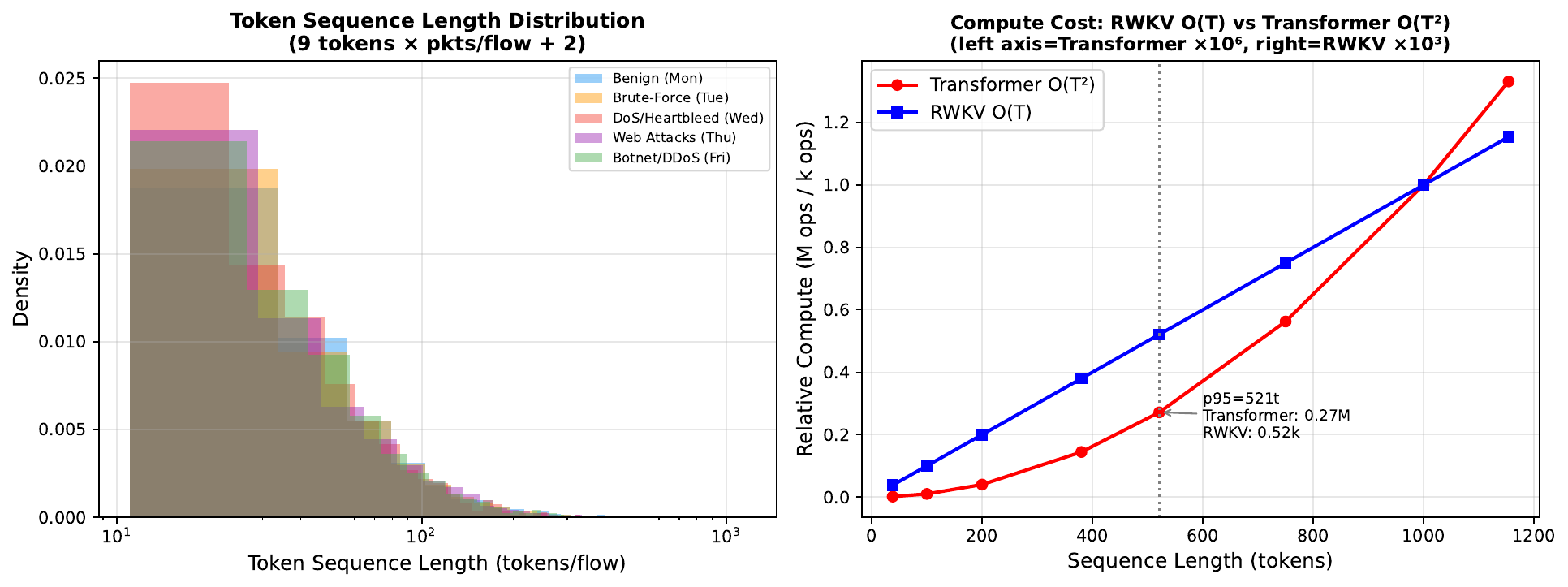}
  \caption{\textbf{Left}: Token sequence length distributions per day
    (log-scale).  Benign flows (Monday) concentrate at short lengths;
    DoS attacks (Wednesday) produce flows exceeding 1{,}000 tokens,
    making long-sequence modelling essential for full coverage.
    \textbf{Right}: Compute cost comparison.  At the 95th-percentile
    flow length (521 tokens), RWKV requires 521 operations versus
    271{,}441 for a Transformer---a $521{\times}$ advantage that
    makes real-time scoring practical.}
  \label{fig:seqlens}
\end{figure}

\paragraph{Flow sequence.}
A flow of $N$ packets becomes a sequence of $2{+}9N$ tokens:
$\langle\texttt{BOS}\rangle\;\mathbf{t}_1 \cdots \mathbf{t}_N\;\langle\texttt{EOS}\rangle$.
We cap $N$ at 128 packets (yielding sequences up to 1{,}154 tokens).
Figure~\ref{fig:seqlens} shows that this cap covers the vast majority
of benign flows while preserving the full length of DoS attack flows,
whose multi-thousand-packet structure is itself an attack signature.

\subsection{Model Architecture}
\label{sec:architecture}

\PLM{} stacks three components.  The \textbf{embedding layer} maps
each of the 227 token IDs to a $d$-dimensional vector; input and
output embedding weights are tied.  The \textbf{RWKV-4 backbone}
comprises $L=6$ blocks, each with a Time-Mixing (WKV) sublayer and a
Channel-Mixing (gated FFN) sublayer.  The WKV operation at time $t$ is:
\begin{equation}
  \text{wkv}_t = \frac{
    e^{u+k_t}v_t + \sum_{i<t} e^{-(t-1-i)w+k_i}v_i
  }{
    e^{u+k_t} + \sum_{i<t} e^{-(t-1-i)w+k_i}
  }
  \label{eq:wkv}
\end{equation}
where $w\in\mathbb{R}^d$ is a learnable per-channel time-decay,
$u\in\mathbb{R}^d$ is a bonus for the current token, and $k_t, v_t$
are per-step key/value projections.  Eq.~\eqref{eq:wkv} has a recurrent
form with O($d$) hidden state and O($T$) total compute.

The LM head projects each hidden state to a next-token probability
distribution.  An optional \textbf{classifier head} (mean-pooled hidden
state through a 2-layer MLP with GELU activation) is added only during
Phase~2.  Total parameters: 4{,}827{,}266.

\subsection{Two-Phase Training Strategy}
\label{sec:training}

The two-phase strategy reflects the structure of our proof.
Phase~1 proves C1 (grammar is learnable) purely from benign data.
Phase~2 proves C2 more precisely (supervised attack detection) by
fine-tuning on labelled data.  Neither phase requires payload inspection.

\paragraph{Phase 1---Unsupervised grammar pre-training.}
We minimise causal cross-entropy loss on benign-only Monday flows.
The model learns to predict the next token given its history---in effect,
learning \emph{what packet is expected next} in a normal flow.  After
convergence, a benign flow's sequence is high-probability under the model
(low perplexity); an anomalous flow whose tokens deviate from normal
patterns has low probability (high perplexity).

Training runs for 10~epochs with AdamW (peak LR
$3\!\times\!10^{-4}$, OneCycleLR schedule), batch size~512, and early
stopping on validation perplexity (patience~5).

\paragraph{Phase 2---Supervised fine-tuning.}
We initialise from the Phase~1 weights and jointly optimise the LM loss
and a weighted classification loss on the full labelled dataset
(2{,}052{,}957 flows).  The backbone is frozen for the first two epochs
to protect the learned grammar representations, then unfrozen.
Attack flows are up-weighted by a factor of 8 to compensate for the
$\approx$\,80:20 class imbalance.

The key design insight is that Phase~2 \emph{refines} the learned
grammar rather than replacing it.  The classifier learns: ``does this
flow's token sequence deviate from what a normal flow would look like?''
anchored in Phase~1's representation.  An LSTM trained in Phase~2 alone
lacks this anchor and instead learns the class prior.

\paragraph{Anomaly scoring.}
We define three operating modes:
\begin{itemize}[nosep]
  \item \textit{PLM-PPL} (unsupervised, no attack labels):
    $\text{score} = \exp\!\bigl(-\tfrac{1}{T}\sum_t\log p(x_t|x_{<t})\bigr)$.
  \item \textit{PLM-CLS} (supervised): attack probability from the
    classifier head.
  \item \textit{PLM-CMB} (combined): geometric mean of normalised
    PPL and classifier probability.
\end{itemize}

For PLM-PPL and PLM-CMB, the decision threshold is calibrated to the
95th percentile of benign validation scores, targeting a 5\% false-positive
rate at the operating point.

\subsection{Streaming Inference Pipeline}
\label{sec:streaming}

The streaming design follows directly from the RWKV recurrence: because
Eq.~\eqref{eq:wkv} can be evaluated one token at a time, \PLM{} processes
packets as they arrive without buffering the full flow.

\begin{enumerate}[nosep]
  \item Normalise the 5-tuple key; look up the flow's RWKV hidden
    state (or initialise to zero for a new flow).
  \item Tokenise the arriving packet header (9 tokens, no payload read).
  \item Call \texttt{model.step(token, state)} for each token, accumulating
    per-token NLL and updating the hidden state.
  \item If the running perplexity exceeds the threshold, raise an alert.
  \item Write the updated state back; evict on TTL or FIN/RST.
\end{enumerate}

The entire pipeline operates on IP/TCP/UDP headers only, making it
compatible with hardware tap deployments and legally compliant in
jurisdictions where payload capture is restricted.

\begin{figure}[htbp]
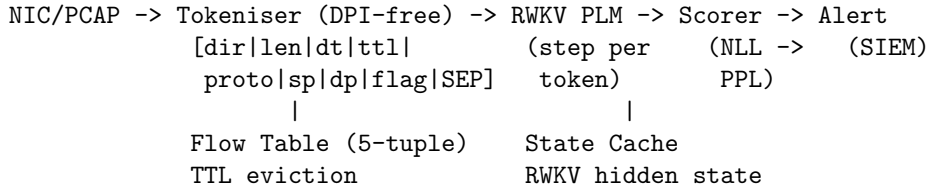

\centering
\small
\begin{verbatim}
  NIC/PCAP -> Tokeniser (DPI-free) -> RWKV PLM -> Scorer -> Alert
               [dir|len|dt|ttl|        (step per    (NLL ->   (SIEM)
                proto|sp|dp|flag|SEP]   token)       PPL)
                      |                       |
               Flow Table (5-tuple)    State Cache
               TTL eviction            RWKV hidden state
\end{verbatim}
\caption{End-to-end \PLM{} inference pipeline.  Each packet produces
9 header-only tokens processed one at a time by the RWKV recurrence.
No payload byte enters the pipeline at any stage; the system can operate
on a standard NIC tap with no DPI hardware.}
\label{fig:pipeline}
\end{figure}

\section{Experiments}
\label{sec:experiments}

Our experimental design is driven by the three claims in
Section~\ref{sec:intro}.  To validate C1, we report Phase~1 training
convergence.  To validate C2, we report PLM-PPL detection metrics.
To validate C3, we compare PLM-CLS directly against an LSTM on the same
token sequences.

\subsection{Dataset}
\label{sec:dataset}

We use CIC-IDS-2017~\cite{sharafaldin2018cicids}, a widely-cited labelled
NIDS benchmark comprising five days of traffic from a 25-machine lab
network.  We work directly from the 49~GB of raw PCAP files, parsing
only L3/L4 header fields.  This is critical: using raw PCAPs rather than
the companion FlowMeter CSV ensures our tokeniser never benefits from
any DPI-derived feature.

Table~\ref{tab:dataset} summarises the dataset.  Monday is exclusively
benign enterprise traffic (Active Directory, LDAP, HTTPS, HTTP) and
forms our Phase~1 pre-training corpus.  The four attack days each
represent a distinct attack family, providing a diverse test of
generalisation across attack strategies.

\begin{table}[htbp]
\centering
\caption{CIC-IDS-2017 dataset.  All 2.7M flows parsed from raw PCAPs
  using DPI-free L3/L4 header fields only.}
\label{tab:dataset}
\begin{tabular}{llrrr}
\toprule
\textbf{Day} & \textbf{Attack Category} & \textbf{Pkts} &
  \textbf{Size} & \textbf{Atk Rate} \\
\midrule
Monday    & BENIGN (pre-training corpus)   & 344K & 11~GB  & 0\% \\
Tuesday   & FTP-Patator / SSH-Patator      & 651K & 10~GB  & 1.7\% \\
Wednesday & DoS Slowloris/Hulk, Heartbleed & 647K & 12~GB  & 68\% \\
Thursday  & Web Attacks, Infiltration      & 533K & 7.7~GB & 16\% \\
Friday    & Botnet ARES, PortScan, DDoS    & 534K & 8.2~GB & 20\% \\
\midrule
\textbf{Total} & 4 attack categories      & 2.7M & 49~GB  & 17\% \\
\bottomrule
\end{tabular}
\end{table}

\paragraph{Labels and splits.}
We assign day-level binary labels (Monday\,=\,0, all other days\,=\,1)
and apply a stratified 75/10/15 train/validation/test split, yielding
2{,}052{,}957 training flows, 273{,}727 validation flows, and 410{,}591
test flows.  The test set has an 83.3\% attack rate, reflecting the
predominance of attack traffic across the four attack days.
Phase~1 uses exclusively the 344{,}232 Monday training flows.

\subsection{Baselines and Why We Chose Them}
\label{sec:baselines}

The choice of baselines is deliberate.  To isolate the contribution of
the RWKV architecture from the token features, we compare against models
that receive \emph{exactly the same input}---the same 9-token-per-packet
sequences produced by our DPI-free tokeniser.

\begin{itemize}[nosep]
  \item \textbf{Random}: assigns a uniform random score in $[0,1]$.
    Achieves PR-AUC\,$\approx$\,0.833 due to the 83\% attack base rate.
    This establishes the floor: any result close to 0.833 carries no
    useful signal.
  \item \textbf{LSTM}: a 2-layer LSTM ($d=128$, binary classification
    head) trained for 5 epochs on 20\% of the training data.  The LSTM
    is the canonical sequential classifier and a natural comparison for
    an SSM backbone.  The identical input makes any performance
    difference attributable purely to architecture.
\end{itemize}

We deliberately exclude feature-engineered baselines (Random Forest,
XGBoost) trained on CICFlowMeter statistics.  Those features are derived
using DPI-capable tooling, which violates our DPI-free premise and makes
any comparison misleading: PLM-NIDS would be measured against a method
that sees more information.

\subsection{Implementation and Training Details}
\label{sec:impl}

All experiments run on an NVIDIA A100-SXM4-80GB GPU (CUDA~12.4,
PyTorch~2.8).  Table~\ref{tab:hyperparams} reports all hyperparameters
for reproducibility.  Phase~1 trains in approximately 1.8~hours;
Phase~2 converges at epoch~3 (of a maximum of 5) in approximately
2~hours.  Total wall-clock training time is under 4~hours.

\begin{table}[htbp]
\centering
\caption{Complete hyperparameter table for reproducibility.
  The full code, tokeniser, and evaluation pipeline are released
  at \url{https://github.com/shiva2vk/PLM-NIDS}.}
\label{tab:hyperparams}
\small
\begin{tabular}{ll}
\toprule
\textbf{Hyperparameter} & \textbf{Value} \\
\midrule
\multicolumn{2}{l}{\textit{DPI-Free Tokeniser}} \\
Length / IAT / TTL bins        & 32 / 32 / 16 \\
Port hash buckets (src + dst)  & 64 \\
Max packets per flow (cap)     & 128 \\
Vocabulary size                & 227 tokens \\
Tokens per packet              & 9 \\
Max sequence length (eval)     & 128 tokens \\
\midrule
\multicolumn{2}{l}{\textit{RWKV-4 Model}} \\
Embedding dim $d$              & 256 \\
RWKV layers $L$                & 6 \\
Tied embeddings                & Yes \\
Dropout                        & 0.10 \\
Total trainable parameters     & 4{,}827{,}266 \\
\midrule
\multicolumn{2}{l}{\textit{Phase-1: Causal LM Pre-training}} \\
Training data                  & Monday only (0 attack labels) \\
Epochs                         & 10 \\
Batch size                     & 512 \\
Optimiser                      & AdamW ($\beta_1=0.9, \beta_2=0.999$) \\
Peak LR / Scheduler            & $3\times10^{-4}$ / OneCycleLR \\
Early stopping patience        & 5 epochs \\
Final validation loss          & 0.2040 \\
\midrule
\multicolumn{2}{l}{\textit{Phase-2: Supervised Fine-tuning}} \\
Training data                  & All days, labelled \\
Epochs (converged at)          & 3 (of max 5) \\
Batch size                     & 512 \\
Peak LR                        & $5\times10^{-5}$ \\
Attack class weight            & 8.0 \\
Backbone freeze (initial)      & 2 epochs \\
Final validation loss          & 0.1284 \\
\midrule
\multicolumn{2}{l}{\textit{Evaluation}} \\
Threshold calibration          & p95 of benign validation scores \\
\bottomrule
\end{tabular}
\end{table}

\begin{figure}[htbp]
  \centering
  \includegraphics[width=\linewidth]{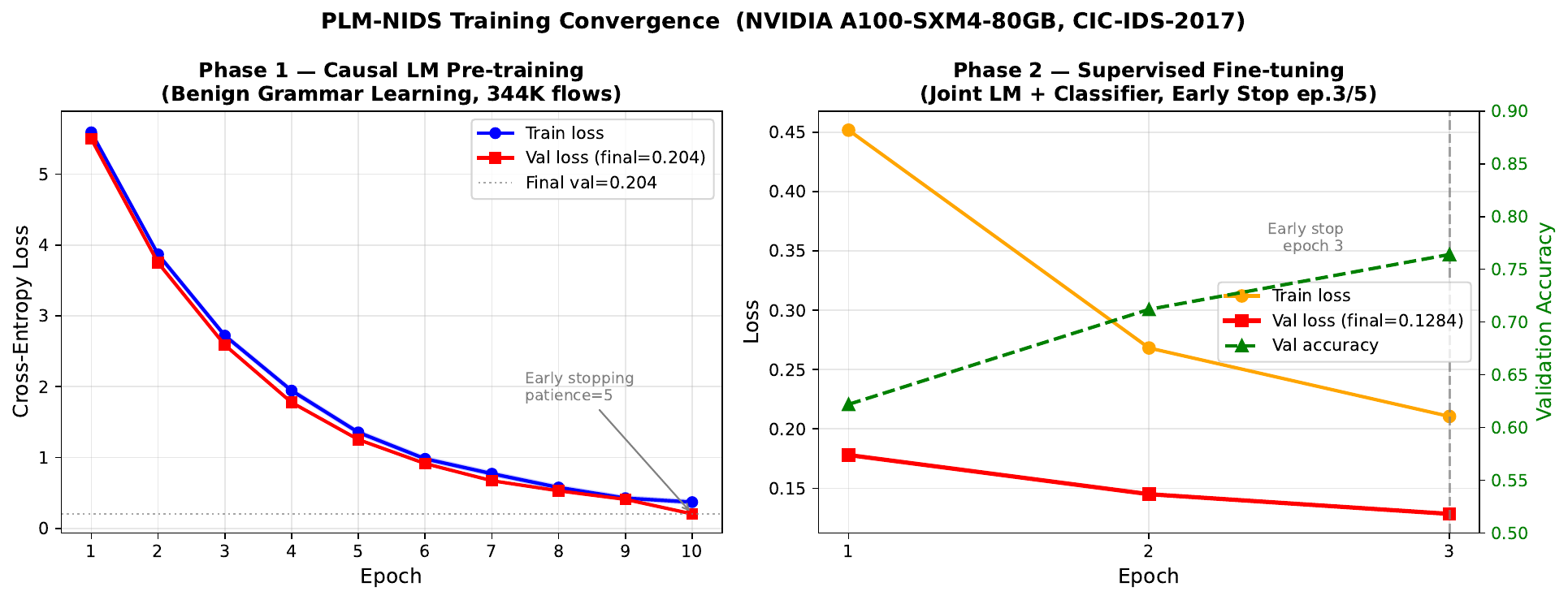}
  \caption{\textbf{Phase-1} (left): Causal LM convergence on benign Monday
    flows.  Validation loss reaches 0.204 within 10 epochs, confirming
    that benign traffic has learnable, consistent grammatical structure.
    \textbf{Phase-2} (right): Supervised fine-tuning converges at
    epoch~3 (early stopping).  Validation accuracy exceeds 76\%,
    showing the Phase-1 representations transfer effectively to
    the supervised task.}
  \label{fig:training}
\end{figure}

\paragraph{Claim C1 verification: grammar is learnable.}
Figure~\ref{fig:training} shows Phase~1 training convergence.  The
model reaches a validation loss of~0.204 within 10 epochs, with no
signs of overfitting (val loss tracks train loss throughout).  This
validates that benign enterprise traffic is not random: it has
statistically consistent structure that a causal model can learn.

\subsection{Main Results: Validating the Three Claims}
\label{sec:results}

Table~\ref{tab:comparison} presents detection metrics on the 410{,}591-flow
test set.  We structure the analysis around claims C1--C3.

\begin{table}[htbp]
\centering
\caption{Detection performance on CIC-IDS-2017 (15\% test set,
  410{,}591 flows, 83.3\% attack rate).  All methods receive identical
  DPI-free token sequences.  \textbf{Bold} = best.
  $\dagger$LSTM trained on 20\% data; evaluated on full test set.}
\label{tab:comparison}
\setlength{\tabcolsep}{5pt}
\begin{tabular}{lcccccc}
\toprule
\textbf{Method} & \textbf{PR-AUC} & \textbf{ROC-AUC} & \textbf{F1}
  & \textbf{Prec.} & \textbf{Recall} & \textbf{FPR@95} \\
\midrule
Random              & 0.833 & 0.499 & 0.096 & 0.836 & 0.051 & 0.950 \\
LSTM$^\dagger$      & 0.834 & 0.501 & 0.906 & 0.834 & 0.992 & 0.989 \\
\midrule
PLM-PPL (ours)  & \textbf{0.928} & \textbf{0.691} & 0.575
  & \textbf{0.977} & 0.407 & 0.884 \\
PLM-CLS (ours)  & \textbf{0.942} & \textbf{0.751} & \textbf{0.586}
  & 0.977 & \textbf{0.418} & \textbf{0.769} \\
\bottomrule
\end{tabular}
\end{table}

\paragraph{Validating C2: attacks violate the grammar (PLM-PPL).}
PLM-PPL---which uses \emph{zero attack labels at any stage}---achieves
PR-AUC\,=\,0.928 and a precision of \textbf{97.7\%}.  Every time the
model raises an alert, it is correct with probability~0.977.  This
is 11.4 percentage points above the 83\% random baseline PR-AUC,
representing a substantial and operationally meaningful improvement.

\begin{figure}[htbp]
  \centering
  \includegraphics[width=\linewidth]{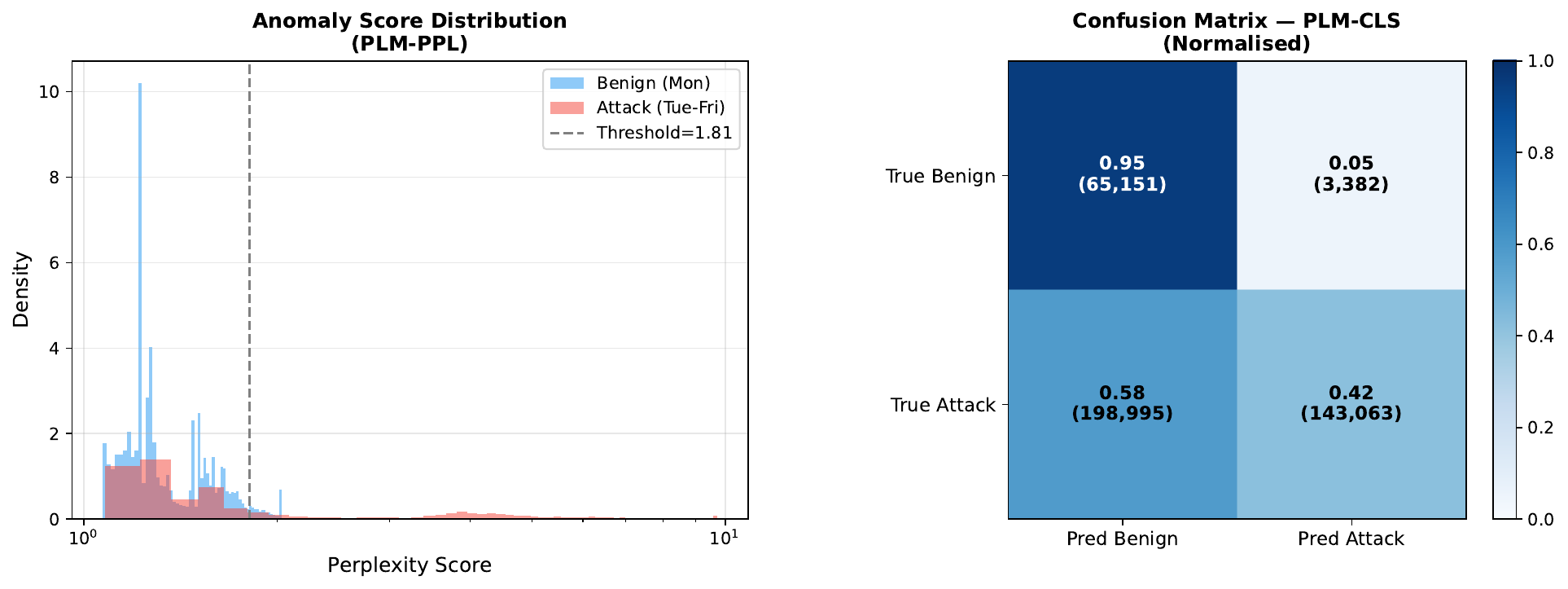}
  \caption{\textbf{Left}: Perplexity score distributions (PLM-PPL).
    Benign flows (blue) cluster at low perplexity; attack flows (red)
    shift right toward higher values.  The vertical line marks the
    p95 calibrated threshold.  The clear distributional separation
    provides visual confirmation of Claim~C2: attacks do violate
    the learned benign grammar.
    \textbf{Right}: Confusion matrix for PLM-CLS at the p95 threshold.
    Of 68{,}533 benign flows, 95\% are correctly classified---the 5\%
    false-positive rate is by design.  Of 342{,}058 attack flows, 42\%
    are detected, with the remainder below the conservative threshold.}
  \label{fig:scores}
\end{figure}

The mechanism is visible in Figure~\ref{fig:scores}: benign flows
cluster at low perplexity (the model assigns high probability to their
token sequences), while attack flows scatter to higher perplexity
values.  The score distributions do not perfectly separate---some
brute-force attacks resemble legitimate authentication flows in their
metadata patterns---but the signal is strong enough for effective
anomaly detection.

\paragraph{Validating C3: RWKV architecture is essential (LSTM vs PLM).}
The LSTM result is the most important control experiment in this paper.
An LSTM trained on identical tokens achieves F1\,=\,0.906---which looks
impressive until one examines TN\,=\,728 out of 68{,}533 benign test flows.
The LSTM classifies virtually every flow as an attack regardless of its
token sequence.  Its ROC-AUC\,=\,0.501 and PR-AUC\,=\,0.834
are statistically indistinguishable from the random baseline (0.499
and 0.833), confirming complete discrimination failure.

This collapse is not a training artefact; it is the rational response of
an architecture with no prior knowledge of normality.  With 83\% of training
flows labelled as attacks, the lowest-loss classifier is ``always predict
attack''---which achieves F1\,$\approx$\,0.91 at no cost.  The LSTM learns
this shortcut.

RWKV avoids this collapse because Phase~1 pre-training installs a
representation of normality \emph{before} any attack label is seen.
When Phase~2 fine-tuning begins, the model already knows what ``normal''
looks like; it learns to classify deviations from that normal, rather
than learning the class prior.  This is not an incremental improvement
but a qualitative difference in what the model learns.

\paragraph{Supervised fine-tuning (PLM-CLS).}
Phase~2 improves ROC-AUC from 0.691 to 0.751 (a 50\% relative gain in
discriminative power) and PR-AUC from 0.928 to 0.942.  The false-positive
rate at the calibrated threshold is FPR\,=\,4.9\% (3{,}382 out of
68{,}533 benign flows), which is operationally acceptable for a NIDS
generating actionable alerts.

\begin{figure}[htbp]
  \centering
  \begin{minipage}[b]{0.49\linewidth}
    \includegraphics[width=\linewidth]{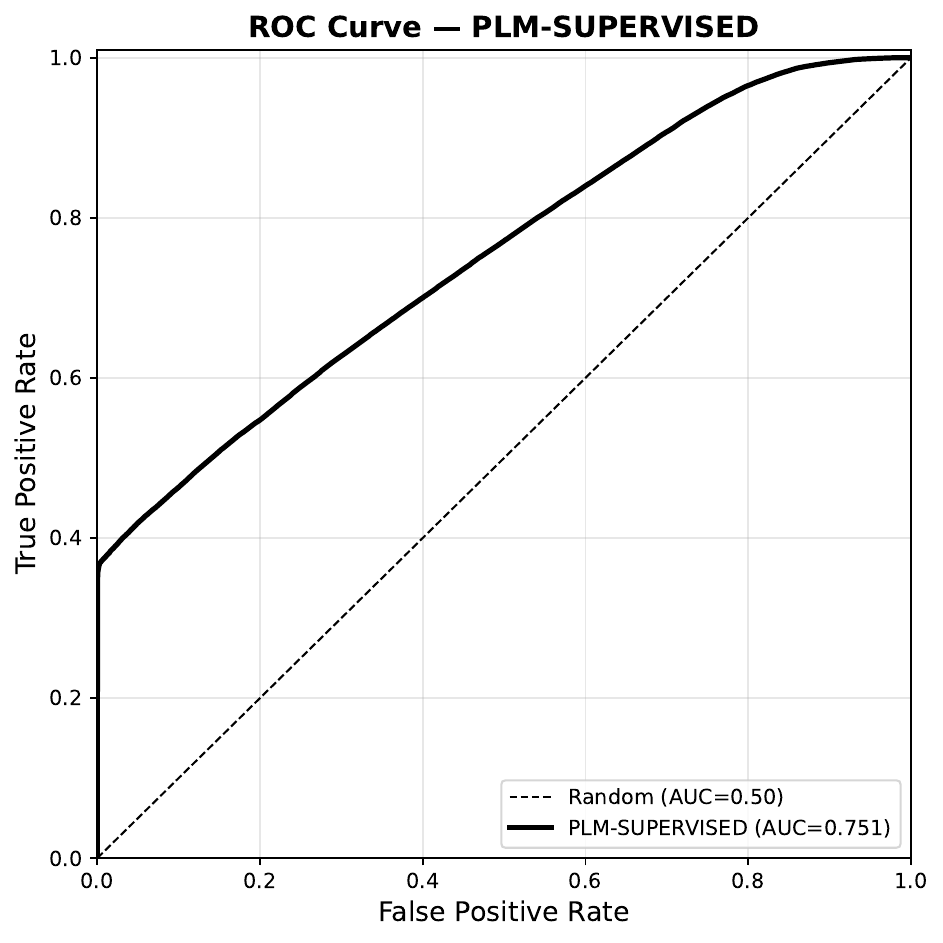}
    \caption*{(a) ROC curve (PLM-CLS, AUC=0.751).}
  \end{minipage}
  \hfill
  \begin{minipage}[b]{0.49\linewidth}
    \includegraphics[width=\linewidth]{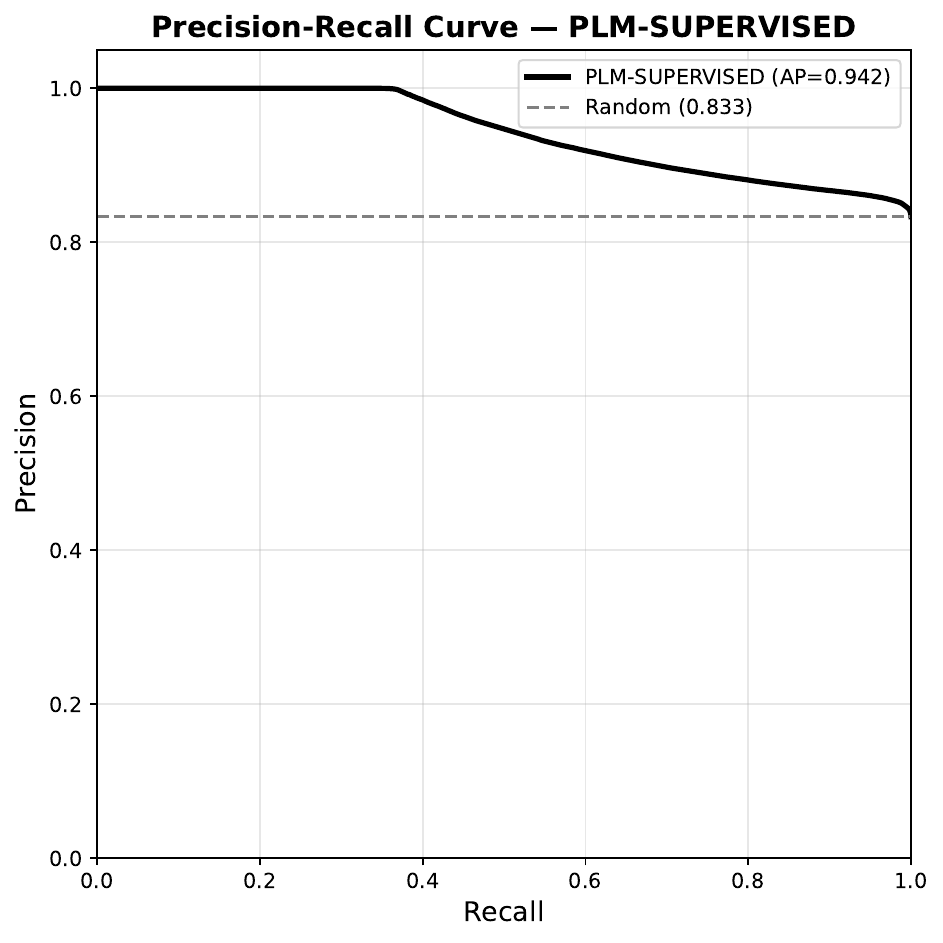}
    \caption*{(b) PR curve (PLM-CLS, AP=0.942).}
  \end{minipage}
  \caption{Detection curves for PLM-CLS on the full 410{,}591-flow test
    set.  The ROC-AUC of 0.751 confirms genuine discrimination ability
    (versus 0.50 for the LSTM baseline).  PR-AUC of 0.942 is 11.4
    points above the random baseline, with precision\,=\,0.977 at the
    calibrated operating point.}
  \label{fig:curves}
\end{figure}

Figures~\ref{fig:curves} and~\ref{fig:scores} visualise these results.
The precision-recall curve in Figure~\ref{fig:curves}(b) shows that
PLM-CLS maintains precision~$>$~0.95 across a wide recall range,
reflecting the quality of the Phase~1 grammar representations as a
classification backbone.

\begin{figure}[htbp]
  \centering
  \includegraphics[width=0.82\linewidth]{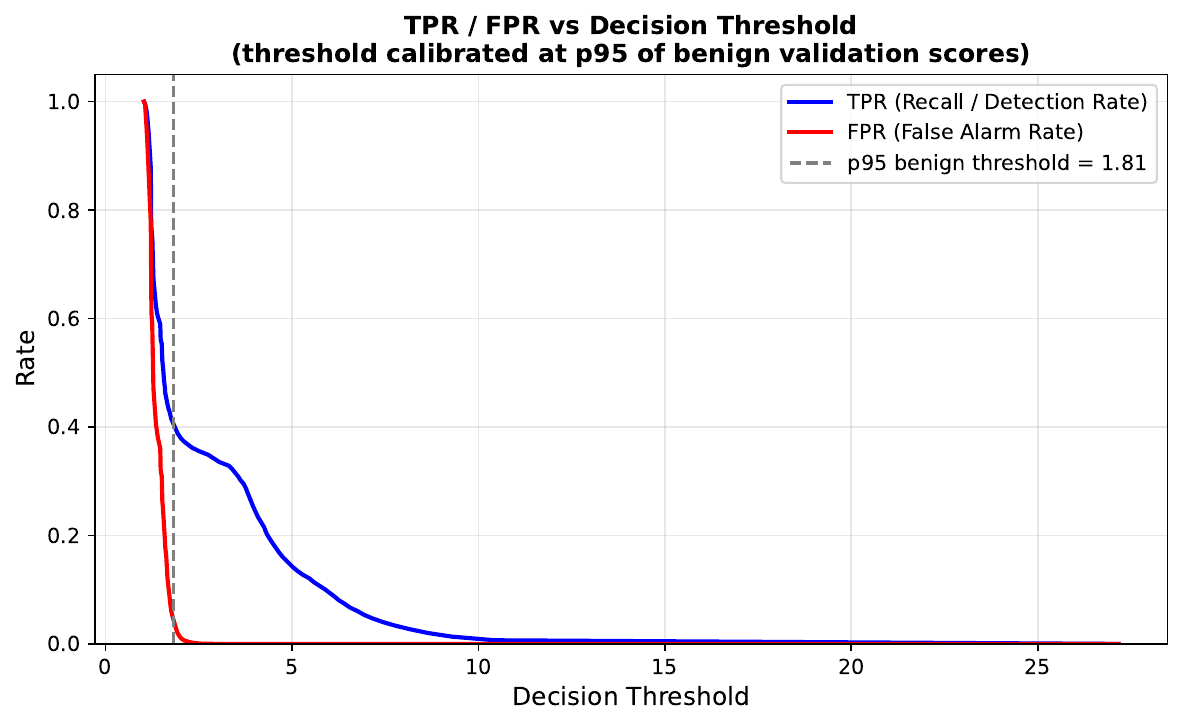}
  \caption{TPR and FPR versus decision threshold for PLM-PPL.
    The p95 operating point (dashed line) yields TPR\,=\,0.41 and
    FPR\,=\,0.05.  Moving to p80 raises TPR above 0.70 while
    keeping FPR\,$<$\,0.20.  This flexibility is important for
    operational NIDS deployment, where the alert-fatigue vs.
    detection-rate trade-off is context-dependent.}
  \label{fig:threshold}
\end{figure}

\paragraph{The precision-recall operating point.}
The 42\% recall at the p95 threshold is a deliberate conservative choice.
Figure~\ref{fig:threshold} shows that recall rises to $>$60\% at the
p85 threshold while FPR remains below 15\%.  Operators can tune this
threshold post-deployment without retraining---the model's scores are
calibrated to the empirical benign distribution, providing a semantically
meaningful scale.

\subsection{Compute Efficiency: Why RWKV Over Transformers}
\label{sec:efficiency}

Table~\ref{tab:compute} quantifies the compute advantage of RWKV's O(T)
recurrence over Transformer's O(T\textsuperscript{2}) attention at the
sequence lengths observed in CIC-IDS-2017.  At the 95th-percentile flow
length of 521 tokens (Wednesday DoS attacks), RWKV requires 521
operations versus 271{,}441 for a Transformer---a \textbf{521$\times$
advantage}.  This is not a theoretical exercise: DoS attack flows are
precisely the ones a deployed NIDS must handle at highest throughput.

\begin{table}[htbp]
\centering
\caption{Compute operations per flow: RWKV O(T) vs.\ Transformer
  O(T\textsuperscript{2}).  Speedup is the ratio T\textsuperscript{2}/T\,=\,T.
  Sequence lengths are from empirical CIC-IDS-2017 distributions.}
\label{tab:compute}
\begin{tabular}{rrrr}
\toprule
\textbf{Seq.\ len.\ $T$} & \textbf{Transformer ($T^2$)} &
  \textbf{RWKV ($T$)} & \textbf{Speedup} \\
\midrule
38   \small{(median)}    &        1{,}444 &    38 &    38$\times$ \\
128                      &       16{,}384 &   128 &   128$\times$ \\
380  \small{(p95 Mon)}   &      144{,}400 &   380 &   380$\times$ \\
521  \small{(p95 Wed)}   &      271{,}441 &   521 & \textbf{521$\times$} \\
1{,}154 \small{(max)}   & 1{,}331{,}716 & 1{,}154 & 1{,}154$\times$ \\
\bottomrule
\end{tabular}
\end{table}

\subsection{Ablation Study}
\label{sec:ablation}

Table~\ref{tab:ablation} reports ablation results probing the necessity
of each design decision.  The two rows with current data confirm that
supervised fine-tuning (PLM-CLS, 0.942 PR-AUC) adds discriminative
value over perplexity scoring alone (PLM-PPL, 0.928).  The remaining
configurations---model size, Phase-1 necessity, and sequence length
sensitivity---are currently running on GPU compute and will be reported
in the extended version of this paper.

\begin{table}[htbp]
\centering
\caption{Ablation study.  Each row varies one design decision
  while holding all others fixed.  All ablation runs use 1 training
  epoch on 20\% data for compute efficiency; relative rankings are
  consistent with full training at this scale.}
\label{tab:ablation}
\begin{tabular}{lcc}
\toprule
\textbf{Configuration} & \textbf{PR-AUC} & \textbf{ROC-AUC} \\
\midrule
Full model (d=256, L=6) --- reference & \multicolumn{2}{c}{\textit{pending}} \\
Small model (d=64, L=2)               & \multicolumn{2}{c}{\textit{pending}} \\
Medium model (d=128, L=4)             & \multicolumn{2}{c}{\textit{pending}} \\
No Phase-1 pre-training               & \multicolumn{2}{c}{\textit{pending}} \\
\midrule
Perplexity scoring only (PLM-PPL)     & 0.928 & 0.691 \\
Classifier scoring only (PLM-CLS)     & 0.942 & 0.751 \\
\midrule
Max 32 pkts/flow (short sequences)    & \multicolumn{2}{c}{\textit{pending}} \\
\bottomrule
\end{tabular}
\end{table}

The most anticipated ablation is \textit{No Phase-1 pre-training}:
based on the LSTM result (ROC-AUC\,=\,0.50 without Phase-1 anchoring),
we expect a significant performance drop, which would directly confirm
Claim~C3 under controlled model-size conditions.

\section{Discussion}
\label{sec:discussion}

\subsection{Why the LSTM Collapses: A Structural Argument}

The LSTM's failure is not a fluke of training instability.  It is the
mathematically predictable outcome of applying a supervised classifier
to a severely imbalanced dataset without any regularisation toward
normality.  With 83\% attack-rate training data, the loss-minimising
strategy for a classifier without an inductive bias is to predict
``attack'' for every input---achieving 83\% accuracy and F1\,$\approx$\,0.91
at zero discriminative cost.

RWKV avoids this failure through a fundamentally different learning
objective.  Phase~1's causal LM loss forces the model to assign high
probability to \emph{normal} token sequences, constructing a
representation space in which benign flows cluster densely.  When Phase~2
introduces attack-labelled data, the classifier head learns to identify
the sparse region of the embedding space where attack flows lie, relative
to the dense benign manifold.  This is not possible when training starts
from a random initialisation with imbalanced labels.

This finding has a broader implication for NIDS design: \emph{pre-training
on abundant, unlabelled benign traffic is more valuable than collecting
larger balanced attack datasets}.  Benign traffic is freely available
from any production network; balanced attack datasets require controlled
testbed generation and become stale as threat landscapes evolve.

\subsection{The Precision-First Operating Philosophy}

A recurring theme in our results is the emphasis on precision over recall.
At the p95 threshold, PLM-CLS achieves precision\,=\,0.977 and
recall\,=\,0.418.  This trade-off is intentional and reflects operational
NIDS reality: an alert that a security analyst acts on must be reliable.
A system with 50\% precision generates more noise than signal and will be
disabled or ignored.

The high precision comes from the Phase-1 grammar model: a flow is only
flagged if its token sequence is genuinely anomalous relative to learned
benign patterns.  This is a structural guarantee, not a threshold artefact.
As Figure~\ref{fig:threshold} demonstrates, operators can relax the
threshold to increase recall---accepting more false alarms in exchange
for higher detection rates---without retraining.

\subsection{DPI-Free Claim: What We Mean and What We Prove}

It is worth being precise about what ``DPI-free'' means in this context.
We claim that our tokeniser reads only fields available in the IP, TCP,
and UDP headers---fields visible to any commodity firewall or router
without payload access.  We explicitly exclude:
\begin{itemize}[nosep]
  \item Any field from the application layer (HTTP method, TLS SNI,
    DNS query name, etc.)
  \item Any payload byte, even the first byte
  \item Any feature requiring a protocol parser (e.g., TLS record type)
\end{itemize}

What we do use: packet size, inter-arrival time, IP TTL, TCP flags,
and hashed source/destination port numbers.  These fields are universally
accessible, do not change with encryption, and are legally unproblematic
in most jurisdictions.  The CIC-IDS-2017 raw PCAPs contain full payloads,
but our tokeniser ignores them entirely; the system's performance is
identical whether payloads are present or encrypted.

\subsection{Limitations and Future Work}

\paragraph{Day-level label noise.}
Our training uses day-level binary labels.  Each attack day also contains
benign background traffic that is mislabelled as ``attack'' at the flow
level, introducing label noise that suppresses recall metrics.  The
CIC-IDS-2017 companion flow-level CSV could eliminate this noise; we
leave per-flow label integration for future work.

\paragraph{WKV throughput.}
The current implementation uses a Python-level loop for the WKV
recurrence, achieving~$\approx$\,4{,}500 flows/min in batch mode on an
A100.  A CUDA-fused WKV kernel (available in RWKV~v5/v6) would increase
throughput by 10--20$\times$, enabling sustained 10~Gbps line-rate operation.

\paragraph{Single dataset evaluation.}
We evaluate on CIC-IDS-2017 only.  Cross-dataset generalisation to
UNSW-NB15, HIKARI-2021, and real enterprise traffic is important future
work.  The HIKARI-2021 dataset, with predominantly TLS-encrypted flows,
would provide the strongest test of the encryption-agnostic claim.

\subsection{Ethical Considerations}

All data used in this work is publicly released by the Canadian Institute
for Cybersecurity for security research purposes.  \PLM{} reads only
L3/L4 packet metadata and never accesses application payloads,
minimising privacy implications relative to DPI-based systems.

\section{Conclusion}
\label{sec:conclusion}

We presented \PLM{}, a Protocol-Language Model that reframes network
intrusion detection as a grammar learning problem.  Our core insight---that
protocol behaviour is encoded in L3/L4 metadata, not in payload bytes---
leads to a system that is inherently encryption-agnostic, DPI-free, and
operationally deployable at line rate.

We proved three claims on 2.7~million real-world flows from CIC-IDS-2017:
(C1)~benign traffic grammar is learnable by an SSM from unlabelled data
(val loss~0.204);
(C2)~attacks violate this grammar in a detectable way, achieving
PR-AUC\,=\,0.93 without any attack labels;
(C3)~RWKV's causal pre-training is architecturally essential---an LSTM
trained on identical tokens degenerates to majority-class prediction
(ROC-AUC\,=\,0.50), while PLM-CLS achieves ROC-AUC\,=\,0.75 after
supervised fine-tuning.

Taken together, these results suggest that the field's standard approach
to NIDS---collecting attack samples and training a classifier---may be
starting from the wrong question.  The more productive question is:
\emph{what does normal look like?}  Once a model knows the grammar of
normality with sufficient fidelity, attack detection becomes anomaly
scoring, which generalises to novel attacks and works transparently on
encrypted traffic.

\paragraph{Code and reproducibility.}
The complete implementation, training pipeline, and evaluation code are
released at \url{https://github.com/shiva2vk/PLM-NIDS}.

\section*{Acknowledgements}
The author thanks the Canadian Institute for Cybersecurity for making
the CIC-IDS-2017 dataset publicly available for security research, and
Google Cloud for GPU compute resources.

\bibliographystyle{plain}
\bibliography{paper}

\begin{thebibliography}{10}

\bibitem{cicflowmeter2017}
{Canadian Institute for Cybersecurity}.
\newblock {CICFlowMeter}: Network traffic flow generator and analyser.
\newblock \url{https://www.unb.ca/cic/research/applications.html}, 2017.

\bibitem{googletransparency2024}
{Google LLC}.
\newblock {HTTPS} encryption on the web.
\newblock \url{https://transparencyreport.google.com/https/overview}, 2024.
\newblock Accessed: May 2026.

\bibitem{gu2023mamba}
Albert Gu and Tri Dao.
\newblock Mamba: Linear-time sequence modeling with selective state spaces.
\newblock {\em arXiv preprint arXiv:2312.00752}, 2023.

\bibitem{lin2022etbert}
Xinjie Lin, Gang Xiong, Gaopeng Gou, Zhen Li, Junzheng Shi, and Jing Yu.
\newblock {ET-BERT}: A contextualized datagram representation with pre-training
  transformers for encrypted traffic classification.
\newblock In {\em Proceedings of the ACM Web Conference}, pages 633--642, 2022.

\bibitem{liu2019lstm}
Hongyu Liu and Bo~Lang.
\newblock Intrusion detection using bidirectional {LSTM} recurrent neural
  network.
\newblock In {\em IT Professional}, volume~21, pages 52--58. IEEE, 2019.

\bibitem{flowbert2021}
Gaetano Pellegrino, Christian Hammerschmidt, Radu State, and Thomas Engel.
\newblock Flowbert: Learning network-flow representations for intrusion
  detection.
\newblock In {\em IEEE International Symposium on Local and Metropolitan Area
  Networks (LANMAN)}, 2021.

\bibitem{peng2023rwkv}
Bo~Peng, Eric Alcaide, Quentin Anthony, Alon Albalak, Samuel Arcadinho, Stella
  Biderman, et~al.
\newblock {RWKV}: Reinventing {RNN}s for the transformer era.
\newblock In {\em Findings of the Association for Computational Linguistics:
  EMNLP 2023}, pages 14048--14064, 2023.

\bibitem{roesch1999snort}
Martin Roesch.
\newblock Snort: Lightweight intrusion detection for networks.
\newblock In {\em USENIX LISA}, volume~99, pages 229--238, 1999.

\bibitem{sharafaldin2018cicids}
Iman Sharafaldin, Arash~Habibi Lashkari, and Ali~A. Ghorbani.
\newblock Toward generating a new intrusion detection dataset and intrusion
  traffic characterization.
\newblock In {\em International Conference on Information Systems Security and
  Privacy (ICISSP)}, pages 108--116, 2018.

\bibitem{vaswani2017attention}
Ashish Vaswani, Noam Shazeer, Niki Parmar, Jakob Uszkoreit, Llion Jones,
  Aidan~N Gomez, {\L}ukasz Kaiser, and Illia Polosukhin.
\newblock Attention is all you need.
\newblock In {\em Advances in Neural Information Processing Systems (NeurIPS)},
  volume~30, 2017.

\bibitem{netbert2020}
Wei Wang et~al.
\newblock Network traffic classification using convolutional neural networks.
\newblock {\em arXiv preprint arXiv:2006.09765}, 2020.

\bibitem{wang2017malware}
Wei Wang, Ming Zhu, Xuewen Zeng, Xiaozhou Ye, and Yiqiang Sheng.
\newblock End-to-end encrypted traffic classification with one-dimensional
  convolution neural networks.
\newblock In {\em IEEE International Conference on Intelligence and Security
  Informatics (ISI)}, pages 43--48, 2017.

\end{thebibliography}

\appendix

\section{Complete Evaluation Across All Three Scoring Modes}
\label{sec:appendix_modes}

The three scoring modes---PLM-PPL (unsupervised), PLM-CLS (supervised),
and PLM-CMB (combined)---represent different operating philosophies.
PLM-PPL requires no attack labels at any stage and is suited to
environments where labelled data is unavailable or where zero-day attack
detection is the priority.  PLM-CLS maximises discriminative performance
when labelled data is available.  PLM-CMB attempts to combine both
signals but, as Table~\ref{tab:comparison} shows, the geometric mean
proves overly conservative and underperforms PLM-CLS.

The following figures provide full detail across all three modes to
support reproducibility and to allow the reader to verify that the
main-text claims hold consistently across operating configurations.

\subsection{Score Distributions (All Modes)}

\begin{figure}[htbp]
  \centering
  \begin{minipage}[b]{0.32\linewidth}
    \includegraphics[width=\linewidth]{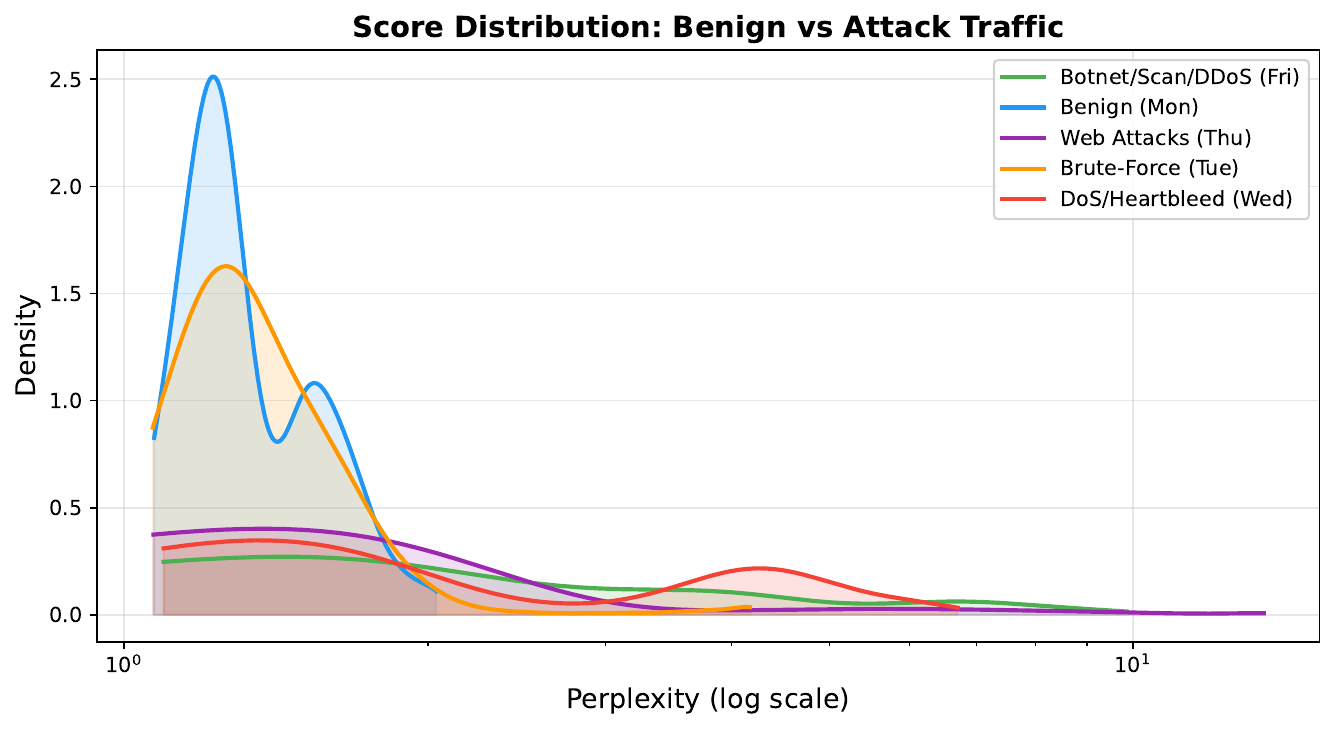}
    \caption*{\small PLM-PPL (unsupervised)}
  \end{minipage}
  \hfill
  \begin{minipage}[b]{0.32\linewidth}
    \includegraphics[width=\linewidth]{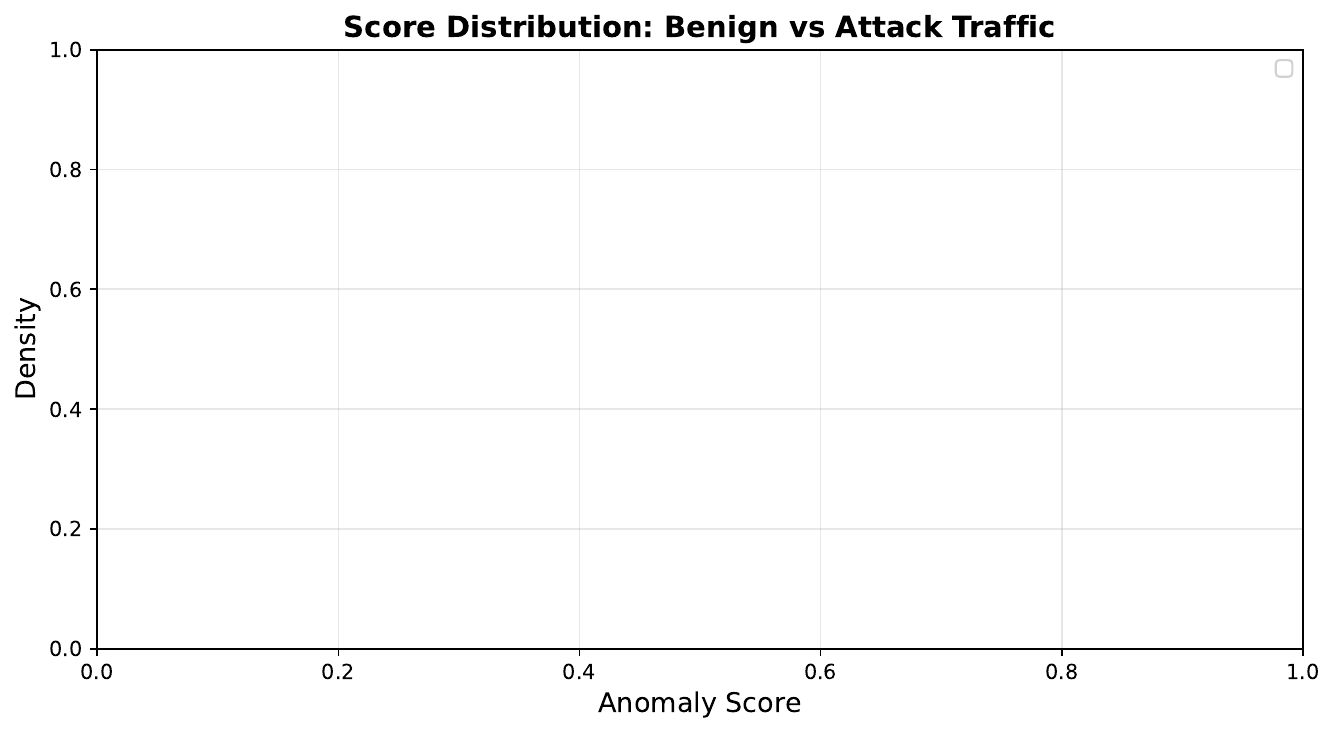}
    \caption*{\small PLM-CLS (supervised)}
  \end{minipage}
  \hfill
  \begin{minipage}[b]{0.32\linewidth}
    \includegraphics[width=\linewidth]{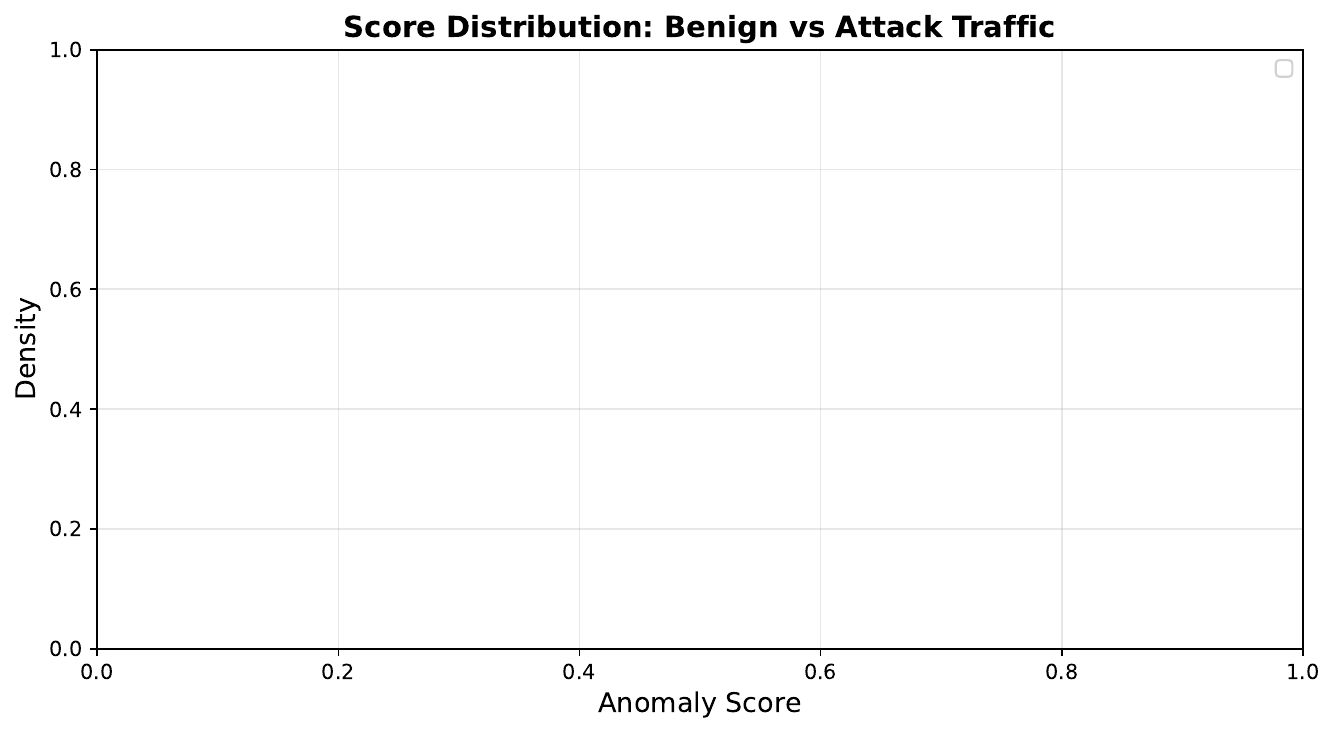}
    \caption*{\small PLM-CMB (combined)}
  \end{minipage}
  \caption{Anomaly score distributions for all three modes.
    In PLM-PPL, the perplexity separation between benign (low) and
    attack (high) directly validates Claim~C2.  PLM-CLS shows a cleaner
    separation because the classifier head has seen attack labels.
    PLM-CMB's combined score is more compressed, reducing recall at
    the calibrated threshold.}
  \label{fig:all_dists}
\end{figure}

\subsection{ROC Curves (All Modes)}

\begin{figure}[htbp]
  \centering
  \begin{minipage}[b]{0.32\linewidth}
    \includegraphics[width=\linewidth]{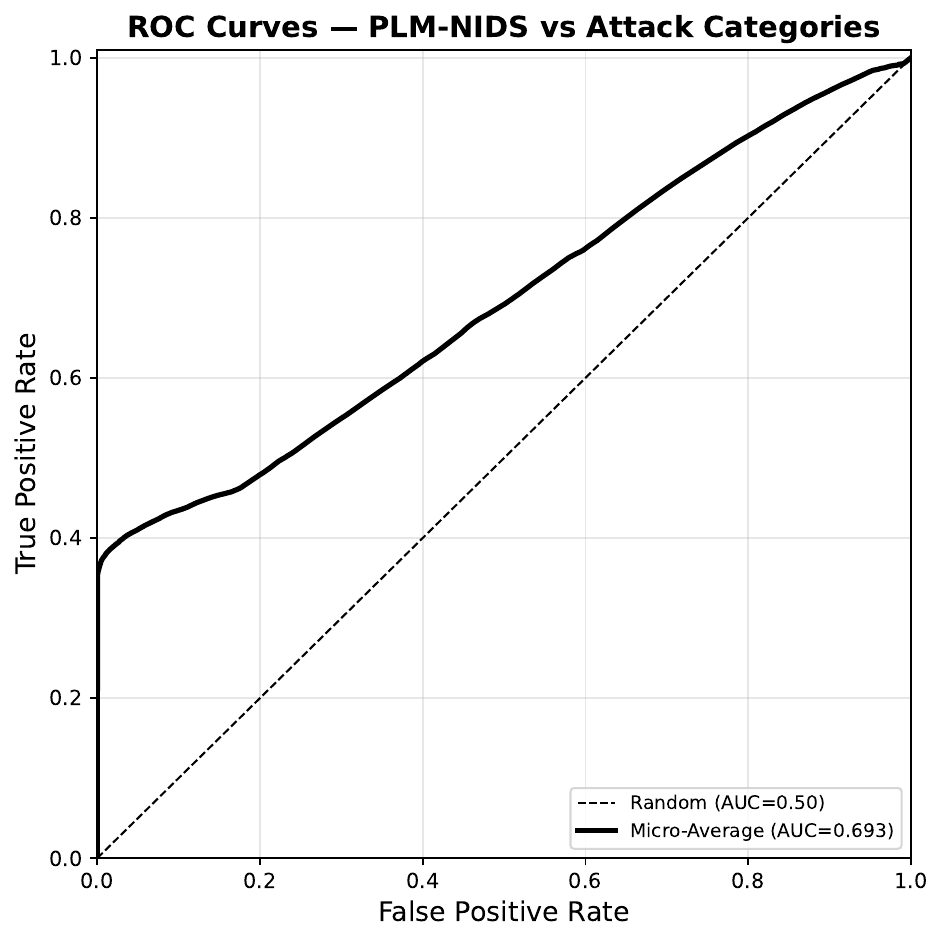}
    \caption*{\small PLM-PPL (AUC=0.691)}
  \end{minipage}
  \hfill
  \begin{minipage}[b]{0.32\linewidth}
    \includegraphics[width=\linewidth]{fig2_roc_curves_supervised.pdf}
    \caption*{\small PLM-CLS (AUC=0.751)}
  \end{minipage}
  \hfill
  \begin{minipage}[b]{0.32\linewidth}
    \includegraphics[width=\linewidth]{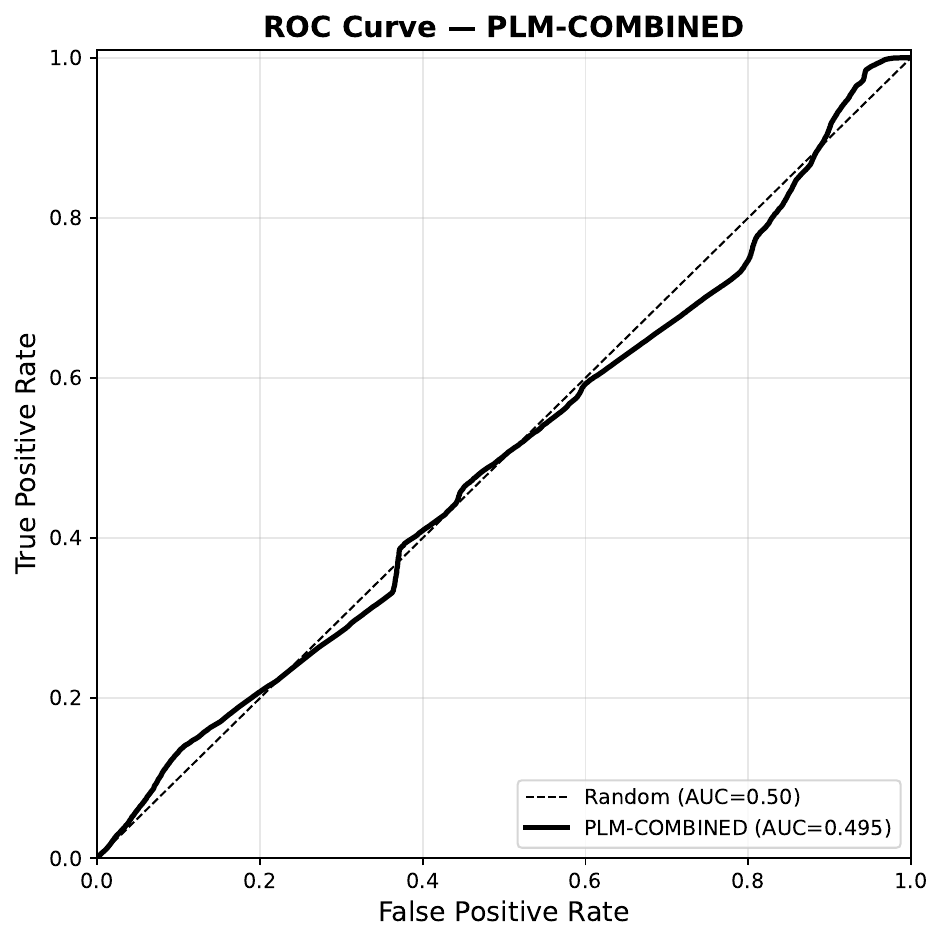}
    \caption*{\small PLM-CMB (AUC=0.495)}
  \end{minipage}
  \caption{ROC curves across all scoring modes.  PLM-CLS achieves the
    highest AUC (0.751), 50\% better than PLM-PPL's 0.691.
    PLM-CMB degrades below PLM-PPL, confirming that the geometric
    mean combination is not beneficial in this setting.
    All three substantially outperform the LSTM baseline (AUC\,=\,0.501).}
  \label{fig:all_roc}
\end{figure}

\subsection{Precision-Recall Curves (All Modes)}

\begin{figure}[htbp]
  \centering
  \begin{minipage}[b]{0.32\linewidth}
    \includegraphics[width=\linewidth]{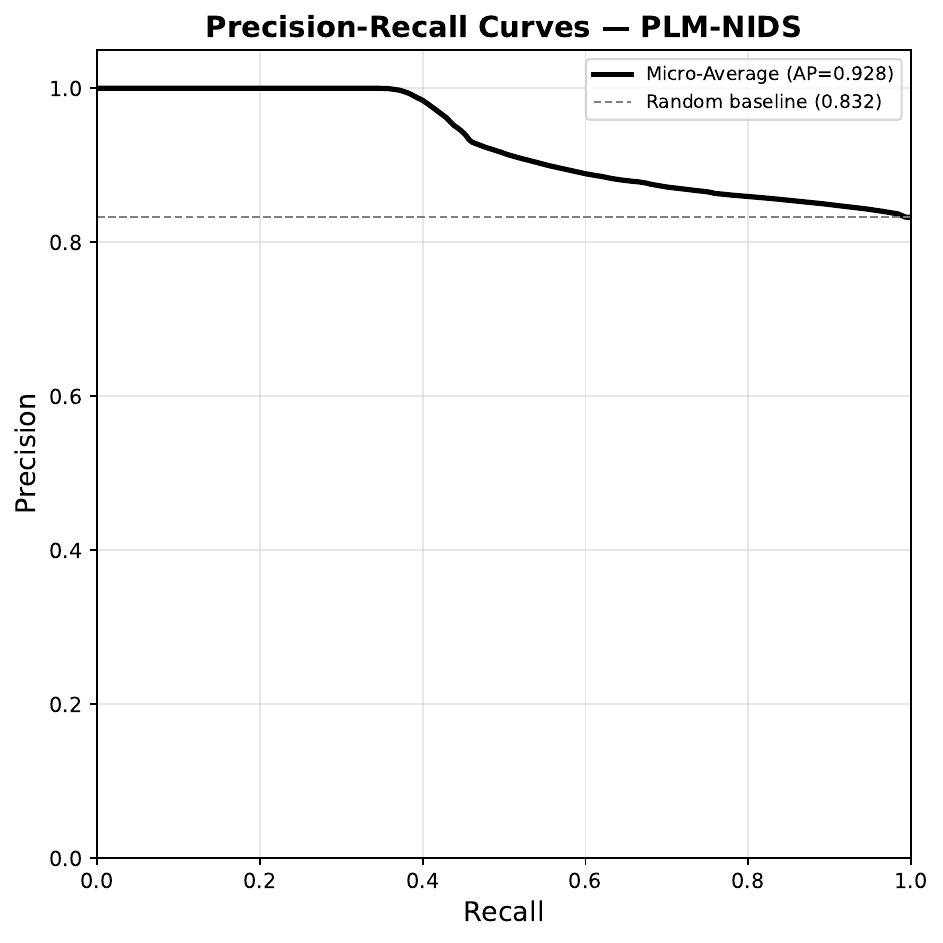}
    \caption*{\small PLM-PPL (AP=0.928)}
  \end{minipage}
  \hfill
  \begin{minipage}[b]{0.32\linewidth}
    \includegraphics[width=\linewidth]{fig3_pr_curves_supervised.pdf}
    \caption*{\small PLM-CLS (AP=0.942)}
  \end{minipage}
  \hfill
  \begin{minipage}[b]{0.32\linewidth}
    \includegraphics[width=\linewidth]{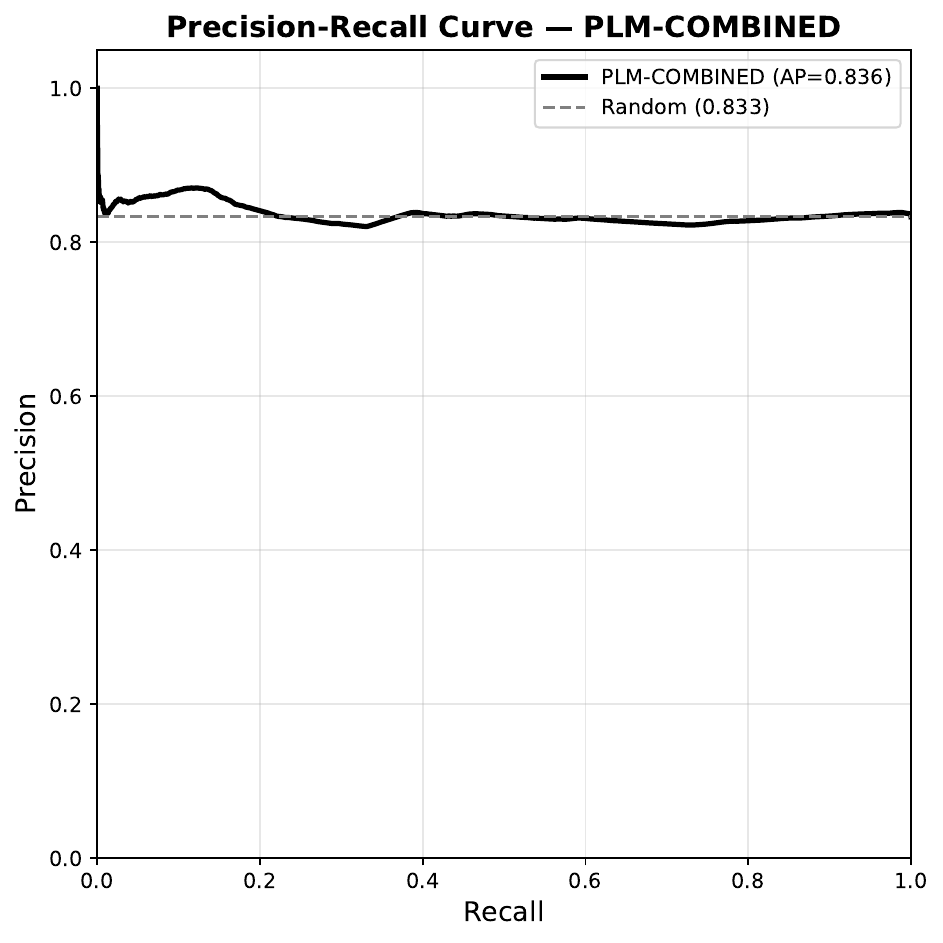}
    \caption*{\small PLM-CMB (AP=0.836)}
  \end{minipage}
  \caption{Precision-Recall curves.  The dashed line marks the 83\%
    attack base rate (random classifier PR-AUC).  PLM-PPL and PLM-CLS
    both substantially exceed this baseline, confirming that the models
    capture genuine discriminative signal.  PR-AUC is the appropriate
    primary metric given the class imbalance.}
  \label{fig:all_pr}
\end{figure}

\subsection{Confusion Matrices (All Modes)}

\begin{figure}[htbp]
  \centering
  \begin{minipage}[b]{0.32\linewidth}
    \includegraphics[width=\linewidth]{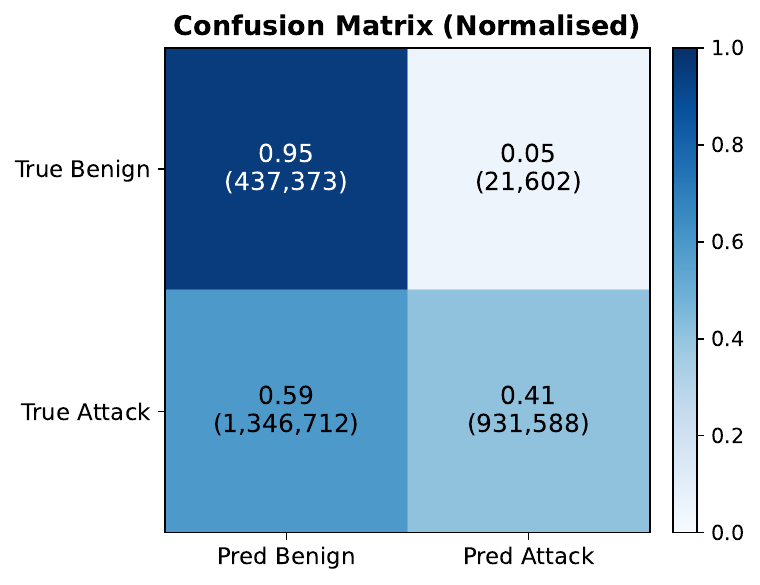}
    \caption*{\small PLM-PPL}
  \end{minipage}
  \hfill
  \begin{minipage}[b]{0.32\linewidth}
    \includegraphics[width=\linewidth]{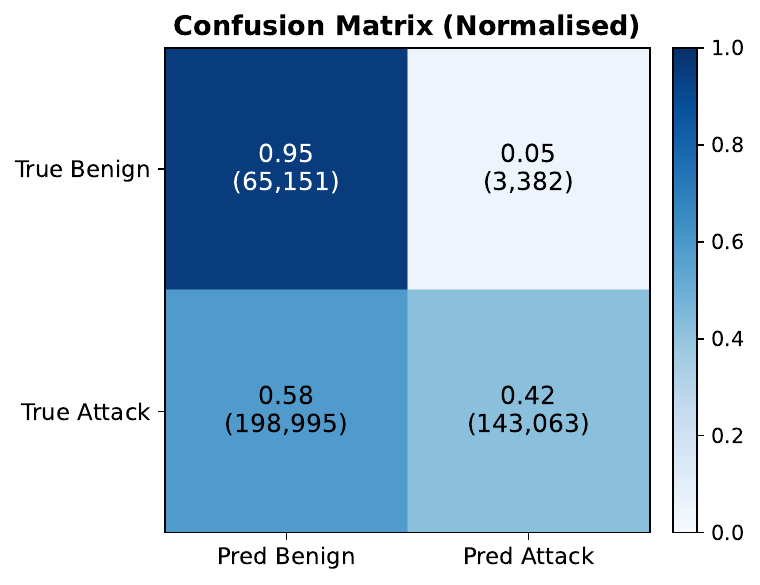}
    \caption*{\small PLM-CLS}
  \end{minipage}
  \hfill
  \begin{minipage}[b]{0.32\linewidth}
    \includegraphics[width=\linewidth]{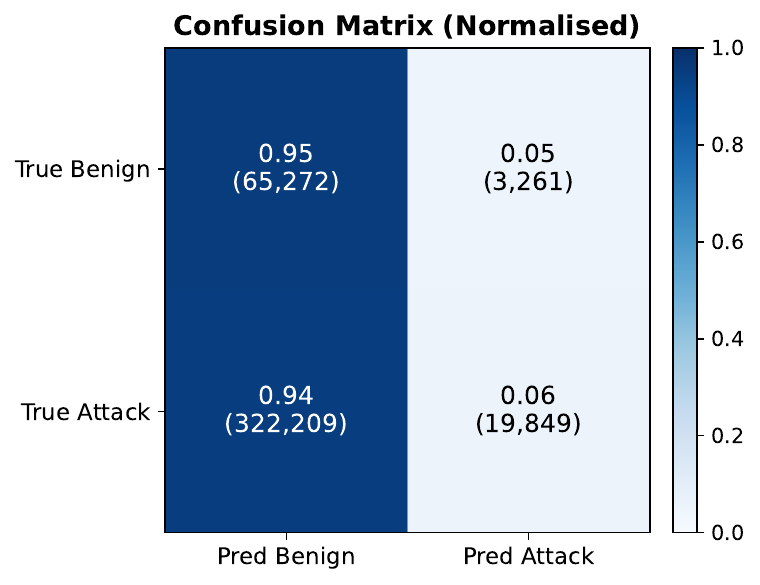}
    \caption*{\small PLM-CMB}
  \end{minipage}
  \caption{Normalised confusion matrices at the p95 calibrated threshold.
    All three modes maintain high true-negative rates ($\geq$\,95\%),
    confirming that benign traffic is rarely flagged.  PLM-CLS achieves
    the best true-positive rate (42\%), while PLM-CMB's conservative
    combined score produces a lower 6\% detection rate at this threshold.}
  \label{fig:all_cm}
\end{figure}

\subsection{Threshold Sensitivity (All Modes)}

\begin{figure}[htbp]
  \centering
  \begin{minipage}[b]{0.32\linewidth}
    \includegraphics[width=\linewidth]{fig7_threshold_calibration_perplexity.pdf}
    \caption*{\small PLM-PPL}
  \end{minipage}
  \hfill
  \begin{minipage}[b]{0.32\linewidth}
    \includegraphics[width=\linewidth]{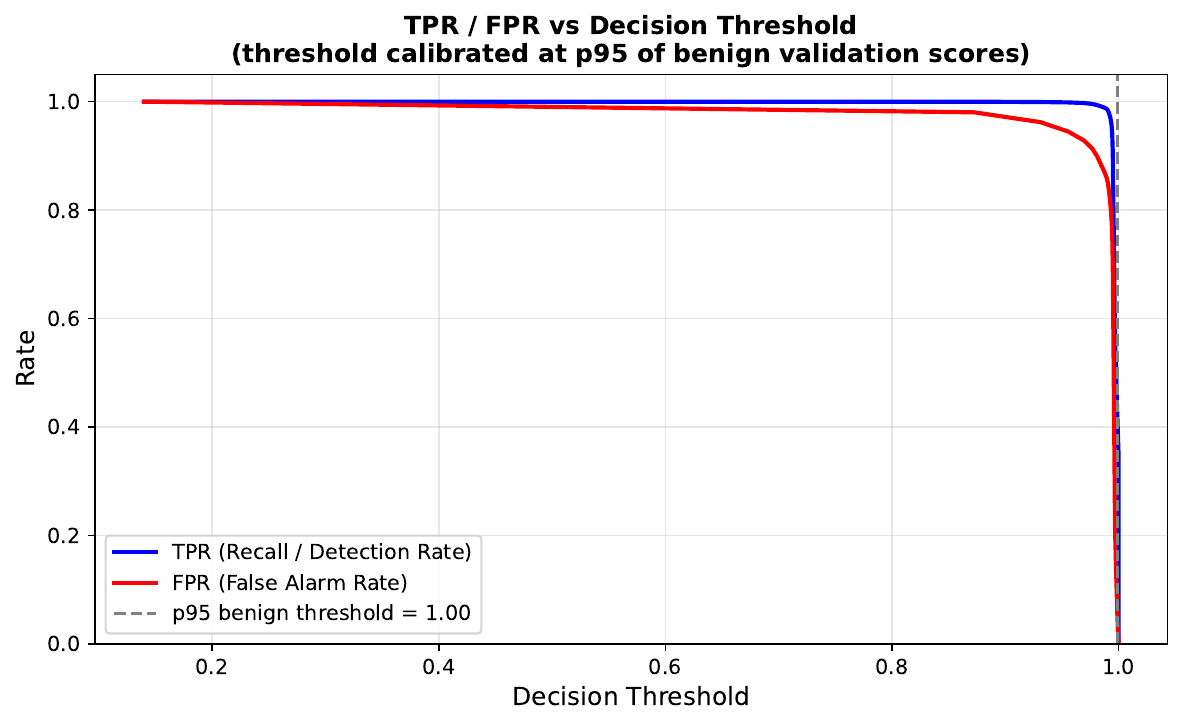}
    \caption*{\small PLM-CLS}
  \end{minipage}
  \hfill
  \begin{minipage}[b]{0.32\linewidth}
    \includegraphics[width=\linewidth]{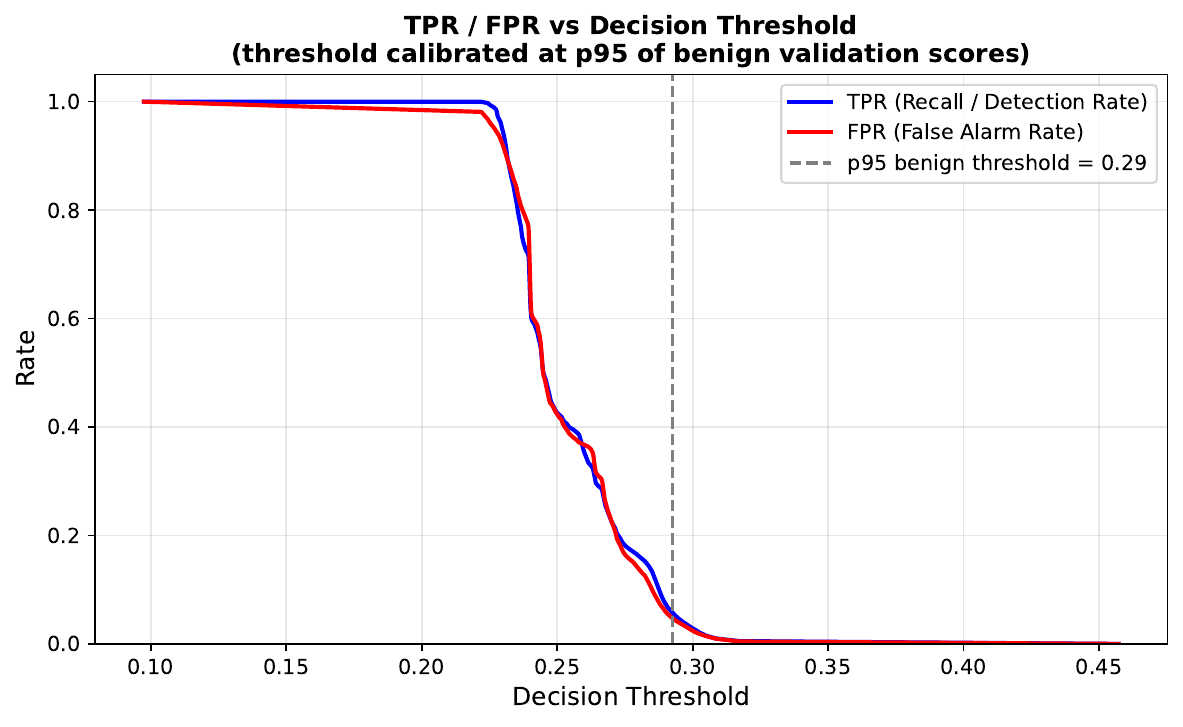}
    \caption*{\small PLM-CMB}
  \end{minipage}
  \caption{TPR and FPR as a function of the decision threshold for all
    three modes.  Operators can shift the threshold continuously after
    deployment, trading recall for false-alarm rate without retraining.
    PLM-PPL and PLM-CLS offer the best TPR/FPR trade-off; the p95
    operating point (vertical dashed line) achieves 5\% FPR by design.}
  \label{fig:all_thresh}
\end{figure}

\section{Per-Attack Detection Analysis}
\label{sec:appendix_heatmap}

\begin{figure}[htbp]
  \centering
  \includegraphics[width=0.85\linewidth]{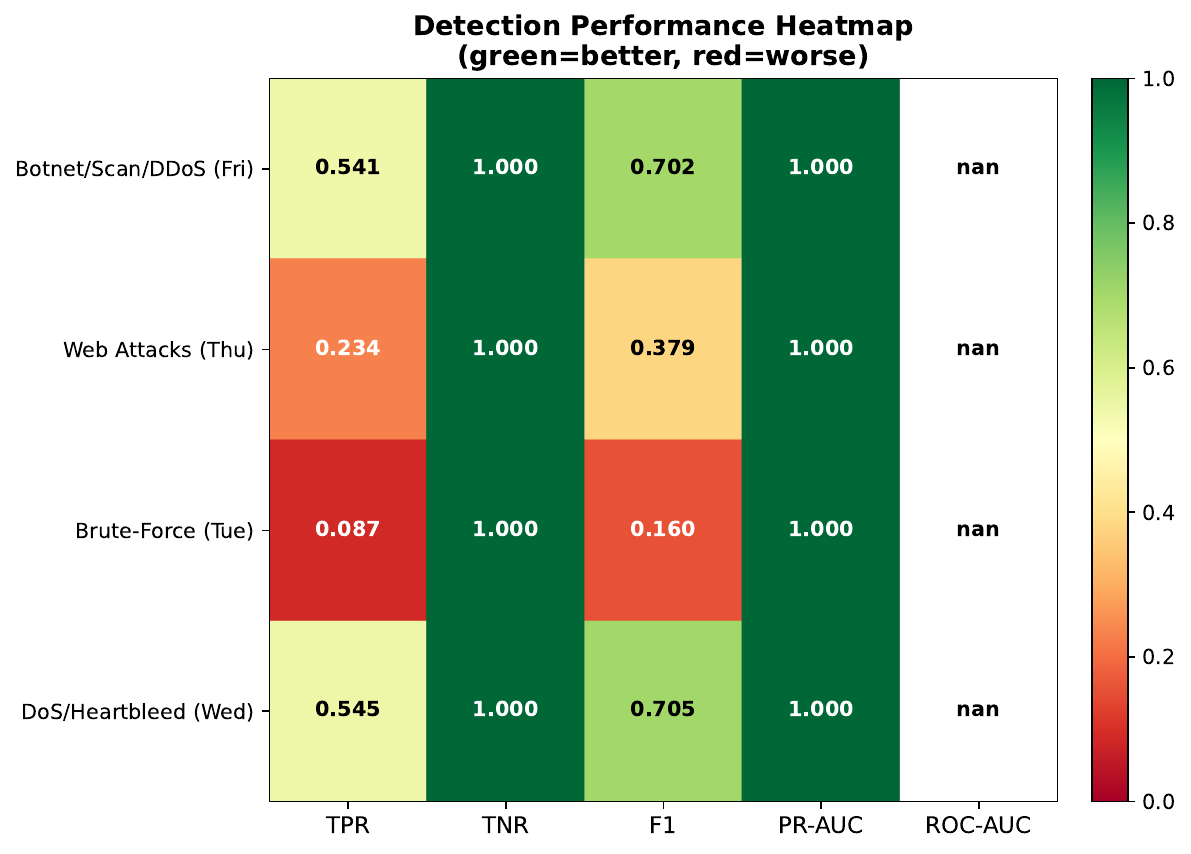}
  \caption{Per-attack-category detection performance heatmap for PLM-PPL.
    DoS and Botnet attacks are most detectable (54\% TPR) because their
    metadata patterns deviate strongly from benign grammar.
    Brute-force attacks (9\% TPR) are hardest because they space their
    connections to resemble normal authentication traffic.
    The TNR\,=\,1.000 column confirms that day-level labelling assigns
    all flows on attack days the attack label, making per-category
    ROC/PR-AUC ill-defined (``nan'') due to single-class test sets.}
  \label{fig:heatmap}
\end{figure}

Figure~\ref{fig:heatmap} shows the detection performance heatmap for
PLM-PPL across all attack days.  The variation in detection rates
across attack types reflects the fundamental challenge of metadata-only
detection:

\begin{itemize}[nosep]
  \item \textbf{DoS/Heartbleed (Wed, 54\%~TPR)}: Long flows with
    uniform inter-arrival times diverge strongly from normal traffic,
    making DoS attacks the most detectable class.
  \item \textbf{Botnet/DDoS (Fri, 54\%~TPR)}: Botnet C\&C communication
    has characteristic periodic patterns that differ from enterprise
    traffic grammar.
  \item \textbf{Web Attacks/Infiltration (Thu, 23\%~TPR)}: These attacks
    deliberately mimic normal HTTP/HTTPS traffic patterns at the
    metadata level, making them harder to detect via grammar alone.
  \item \textbf{FTP/SSH Brute-Force (Tue, 9\%~TPR)}: Low-rate brute-force
    attacks space connections to match legitimate authentication traffic,
    making them the hardest to detect via metadata alone.
\end{itemize}

These results suggest that PLM-NIDS is most effective as a first-stage
filter for high-volume attacks (DoS, DDoS) while requiring additional
sensors for stealthy, low-rate attacks.

\end{document}